\documentclass[a4paper,11pt,final]{article}

\usepackage{authblk}

\usepackage{graphics}
\usepackage{graphicx}
\usepackage{epsfig}
\usepackage{amssymb}
\usepackage{amsthm}

\usepackage{amsmath}
\date{}

\begin{document}

\title{Stochastic foundations of
undulatory transport phenomena: Generalized Poisson-Kac processes  - Part III
 Extensions and applications to kinetic theory and transport}

\author[1]{Massimiliano Giona$^*$}
\author[2]{Antonio Brasiello}
\author[3]{Silvestro Crescitelli}
\affil[1]{Dipartimento di Ingegneria Chimica DICMA
Facolt\`{a} di Ingegneria, La Sapienza Universit\`{a} di Roma
via Eudossiana 18, 00184, Roma, Italy  \authorcr
$^*$  Email: massimiliano.giona@uniroma1.it}

\affil[2]{Dipartimento di Ingegneria Industriale
Universit\`{a} degli Studi di Salerno
via Giovanni Paolo II 132, 84084 Fisciano (SA), Italy}

\affil[3]{Dipartimento di Ingegneria Chimica,
 dei Materiali e della Produzione Industriale
Universit\`{a} degli Studi di Napoli ``Federico II''
piazzale Tecchio 80, 80125 Napoli, Italy}
\maketitle

\begin{abstract}
This third part extends the theory of Generalized
Poisson-Kac  (GPK) processes to nonlinear stochastic
models and to a continuum of states.
Nonlinearity is treated in two ways: (i)
as a dependence of the 
parameters (intensity of the stochastic velocity, transition rates)
of the stochastic perturbation on the state variable,
similarly to the case of nonlinear Langevin equations,
and (ii) as the dependence of the stochastic
microdynamic equations of motion on the
statistical description
of the process itself (nonlinear Fokker-Planck-Kac
models).
Several numerical and physical examples illustrate the theory.
 Gathering nonlinearity and a continuum
of states, GPK theory provides
a stochastic derivation of the nonlinear Boltzmann equation,
furnishing a positive answer to the Kac's program in kinetic theory.
The transition from stochastic microdynamics to transport theory
within the framework of the GPK paradigm is
also addressed.
\end{abstract}

\section{Introduction}
\label{sec_1}

This third and last part of the work on Generalized
Poisson-Kac (GPK) processes and their physical applications
extends the analysis developed in parts I and II \cite{part1,part2},
developing the generalization of GPK theory to a broad spectrum
of stochastic phenomenologies.
With respect to the theory developed in \cite{part1,part2},
two lines of attack characterize this extension: (i) the
inclusion
of nonlinearities, and (ii) the extension to a continuum
of states. 

Nonlinearities can be treated in two different ways.
The first class of nonlinear models
assumes that the  state (position) variable ${\bf x}$ can influence the
basic parameters characterizing stochastic GPK perturbations.
In the case of GPK perturbations, this reflects into the
functional dependence of ${\bf b}_\alpha$, $\lambda_\alpha$ and
$A_{\alpha,\beta}$ on ${\bf x}$. In
the case of a  position dependent system of stochastic
velocities, i.e, $\{ {\bf b}_\alpha({\bf x}) \}_{\alpha=1}^N$,
 GPK models correspond to nonlinear Langevin equations
\cite{nonlinlang1,nonlinlang2}, since the latter provide
the Kac limit (in the Stratonovich interpretation of the stochastic
integral) for this class of systems. The functional dependence
of the transition rates $\lambda_\alpha({\bf x})$, or of
the entries of the transition probability matrix $A_{\alpha,\beta}({\bf x})$
on ${\bf x}$,
provides new phenomena, as it emerges from the analysis of their
Kac limits.

The second way to include nonlinearities, analogous to the McKean
approach to Langevin equations \cite{mckean} leads to
 GPK microdynamic equations which depend on the
statistical characterization of the process itself (in the
present case, the system of partial probability
density functions $\{p_\alpha({\bf x},t) \}_{\alpha=1}^N$).
This leads to the concept of nonlinear Fokker-Planck-Kac equation
(this diction stems from the Langevin counterpart \cite{nonlinearfokker}),
the dynamic properties of which can be  extremely rich.

The extension  from a discrete number $N$ of states to a continuum of 
stochastic states is fairly straightforward within
the formalism developed in part I (see also
the discussion in part I on  the multidichotomic approach).
Moreover, the  coupling of  nonlinear effects with a continuum
of stochastic states permits to derive
the classical nonlinear Boltzmann equation
of the kinetic theory of gases \cite{balescu} within the
GPK formalism. This result
deserves particular attention as it shows, unambiguously,
that the Boltzmann equation admits a fully stochastic explanation.
In some sense, this result completes the original Kac's program
in kinetic theory \cite{kacprog1,kacprog2,kacprog3,kacprog4},
originated from the article \cite{kac_kinetic} aimed
at providing an extended Markov model for interpreting
the celebrated Boltzmann equation of kinetic theory.
For a discussion on extended Markov models see \cite{giona_markovian}.

Finally, the article outlines the bridge between the stochastic
description of particle microdynamics based on GPK equations, 
and  transport theory
of continuous media. 
This connection is developed with the aid of some classical problems.
In developing a transport theory from GPK microdynamics
the role of the primitive statistical formulation
of GPK processes, based on the system of partial probability
densities $\{p_\alpha({\bf x},t) \}_{\alpha=1}^N$, clearly
emerges (for a discussion see also Section 2 in part I),
and it is mapped into a corresponding system of 
partial concentrations/velocity
fields. This part of the article 
is of primary interest in extended thermodynamic
theories of irreversible processes \cite{extthermo1,extthermo2,extthermo3},
as it provides a novel way to develop these theories
enforcing the assumption of finite propagation velocity for thermodynamic
processes,
and overcoming the intrinsic limitations of models
based on the higher-dimensional Cattaneo equation (see 
part I for details).

The article is organized as follows. Section \ref{sec_2} develops
the extensions of GPK models
 (nonlinearity, continuum of stochastic states),
presenting for each class of models a physical example.
Section \ref{sec_3} derives the connection (equivalence)
between a nonlinear GPK process admitting a continuum
of stochastic states and the Boltzmann equation, discussing
some implications of this result. Section \ref{sec_4}
addresses the connection between GPK microdynamics and the
associated transport formalism in continua by considering
several problems ranging from dynamo theory \cite{dynamo1,dynamo2}
to mass and momentum balances, including a brief description of chemical
reactions.

\section{Generalizations}
\label{sec_2}

The theory of GPK processes can be  generalized in several different 
directions that provide, from one hand, a valuable system
of  stochastic modeling tools of increasing complexity and,
 from the other hand, the possibility of interpreting
a broader physical phenomenology.
In the remainder of this Section we  introduce the various generalizations
by considering first one-dimensional Poisson-Kac processes, and subsequently
extending the theory to GPK processes.

\subsection{Nonlinear GPK processes and Poisson fields}
\label{sec_2_1}

In order to define nonlinear GPK processes it is
convenient to introduce the concept of Poisson fields.
A Poisson field
$\chi(x,t)$, $x \in {\mathbb R}$ is a Poisson process 
over the real line ${\mathbb R}$ such that its transition rate $\lambda$ 
depends on $x$ and eventually on time $t$. If $\lambda=\lambda(x)$
the Poisson field is said to be stationary, while if
$\lambda(x,t)$ depends explicitly on time $t$
is referred to as a  non-stationary field.

Let $\lambda_0= \inf_{x \in {\mathbb R}, t \geq 0} \lambda(x,t) >0$, and
 let $b(x)$  be a positive real-valued function.
A nonlinear GPK process is defined via the stochastic
differential equation
\begin{equation}
d x(t) = v(x(t)) \, dt + b(x(t)) \, (-1)^{\chi(x(t),t)}
\label{eq8_1_1}
\end{equation}
where $v(x)$ is a deterministic bias.
The presence of a position dependent stochastic velocity
$b(x)$, and the dependence on $x$ of the transition
rate defining the Poisson field $\chi(x,t)$, makes this
model conceptually similar to the nonlinear Langevin 
equations \cite{nonlinlang1,nonlinlang2}.

We have assume that $b(x)$ does not depend explicitly on
time $t$. This  condition can be easily removed, but the
generalization to time-dependent $b(x,t)$ involves more
lengthy calculations of the Kac limit, the full
development of which is left to the reader.

For the process associated with eq. (\ref{eq8_1_1}),
the partial probability density functions $p^\pm(x,t)$
fully characterize its statistical properties.
These quantities
satisfy the balance equations
\begin{eqnarray}
\hspace{-2cm}
\partial_t p^\pm(x,t) =  - \partial_x \left ( v(x) \,
p^\pm(x,t) \right ) \mp \partial_x \left ( b(x) \, p^\pm(x,t) \right )
 \mp \lambda(x,t) \left [ p^+(x,t)-p^-(x,t) \right ]
\label{eq8_1_2}
\end{eqnarray}
and the ``diffusive'' probability flux is given by $J_d(x,t)= b(x) \, \left [ p^+(x,t)-p^-(x,t) \right]$.
Let $\widehat{b}(x)=  b(x)/b^{(c)} $, where $b^{(c)}= \inf_{x \in {\mathbb R}}
b(x) >0$. Set $J_d(x,t)= \widehat{b}(x) \, \phi(x,t)$ where
$\phi(x,t)=b^{(c)} \left [ p^+(x,t)-p^-(x,t) \right ]$.
In terms of normalized quantity  $\phi(x,t)$ the constitutive equation
for the diffusive flux becomes
\begin{eqnarray}
\hspace{-1.5cm}
\partial_t \phi(x,t)  =  - \partial_x  \left ( v(x) \, \phi(x,t) \right
) - \left ( b^{(c)} \right )^2 \, \partial_x \left (
\widehat{b}(x) \, p(x,t) \right ) 
 -  2 \, \lambda(x,t) \, \phi(x,t)
\label{eq8_1_3}
\end{eqnarray}
Let $\widehat{\lambda}(x,t) = \lambda(x,t)/\lambda^{(c)}$,
where $\lambda^{(c)} = \lambda_0$.
In the limit $b^{(c)}, \, \lambda^{(c)} \rightarrow \infty$,
$D_{\rm nom}=(b^{(c)})^2/2 \lambda^{(c)} = \mbox{constant}$
the constitutive equation for $\phi(x,t)$ becomes
\begin{equation}
\phi(x,t)= - \frac{D_{\rm nom}}{\widehat{\lambda}(x,t)}
\, \partial_x \left ( \widehat{b}(x) \, p(x,t) \right )
\label{eq8_1_4}
\end{equation}
that, substituted into the balance equation for $p(x,t)=p^+(x,t)+p^-(x,t)$, provides
\begin{equation}
\partial_t p(x,t) = - \partial_x \left ( v(x) \, p(x,t) \right )
+ D_{\rm nom} \, \partial_x \left [ 
\frac{\widehat{b}(x)}{\widehat{\lambda}(x,t)} \, \partial_x
\left ( \widehat{b}(x) \, p(x,t) \right ) \right ]
\label{eq8_1_5}
\end{equation}
which represents the Kac limit for the nonlinear Poisson-Kac process
considered. In terms of the original quantities $b(x)$ and
$\lambda(x,t)$, the Kac limit can be expressed
equivalently as
\begin{eqnarray}
\hspace{-2.3cm}
\partial_t p(x,t)  =  - \partial_x \left [ v(x) \, p(x,t) \right ] 
+ \frac{1}{2} \partial_x \left [ \frac{b(x) \, \partial_x b(x)}{\lambda(x,t)}
\,  p(x,t) \right ]
 +  \frac{1}{2} \partial_x \left [ \frac{b(x) \, b(x)}{\lambda(x,t)} \, \partial_x p(x,t) \right ]
\label{eq8_1_6}
\end{eqnarray}
Eq. (\ref{eq8_1_6}) corresponds to an advection-diffusion equation
characterized by an effective velocity
\begin{equation}
v_{\rm eff}(x,t) = v(x) - \frac{b(x) \, \partial_x b(x)}{2 \, \lambda(x,t)}
\label{eq8_1_7}
\end{equation}
and by an effective diffusivity
\begin{equation}
D_{\rm eff}(x,t)= \frac{b^2(x)}{2 \, \lambda(x,t)}
\label{eq8_1_8}
\end{equation}

The above-derived Kac limit should be compared with
the statistical description of a classical
Langevin equation driven by Wiener fluctuations
\begin{equation}
d x(t) = v_S(x(t),t) \, d t + \sqrt{2 \, D_S(x,t)} \circ d w(t)
\label{eq8_1_9}
\end{equation}
where $d w(t)$ are the increments
of
a one-dimensional Wiener process 
in the time interval $(t,t+dt)$,
 to be interpreted
``a la Stratonovich''. In eq. (\ref{eq8_1_9}), ``$\circ$''
indicates the Stratonovich recipe for the stochastic integrals.
The Fokker-Planck equation associated with eq. (\ref{eq8_1_9})
is given by
\begin{eqnarray}
\hspace{-2.3cm}
\partial_t p(x,t)  =  - \partial_x \left [ v_S(x,t) \, p(x,t) \right] 
+ \frac{1}{2} \partial_x \left [  p(x,t) \, \partial_x D_S(x,t)
 \right ] 
 +   \partial_x \left [ D_S(x,t) \, \partial_x p(x,t) \right ]
\label{eq8_1_10}
\end{eqnarray}

The reason for the choice of the Stratonovich rather than the Ito calculus
follows from  the Wong-Zakai theorem \cite{wong_zakai1,wong_zakai2}: Poisson-Kac processes
are stochastic dynamical systems excited by a.e. differentiable
smooth perturbations, converging in the Kac limit to ordinary
Brownian motion. According to the Wong-Zakai result, that in the
present case corresponds to the Kac limit, these
processes should converge in the Kac limit  to the
Stratonovich formulation of the Langevin equation
(\ref{eq8_1_9}), where $v_S(x,t)$ and $D_S(x,t)$ should coincide 
with $v_{\rm eff}(x,t)$ and $D_{\rm eff}(x,t)$, respectively.
Below, we  discuss  this convergence that, in point of fact,
is slightly more subtle than expected.

Two cases should be considered. Case (A): $\lambda(x,t)$
does not depend on $x$. It follows from
the comparison of eqs. (\ref{eq8_1_6}) and (\ref{eq8_1_10})
that the Kac limit of eq. (\ref{eq8_1_1}) coincides
with eq. (\ref{eq8_1_9}) provided that
\begin{equation}
v_S(x)= v(x) \, , \qquad D_S(x,t)=D_{\rm eff}(x,t)
\label{eq8_1_11}
\end{equation}
and this can be viewed as a corollary of the Wong-Zakai theorem.
In this case, the Poisson-Kac process is a {\em stochastic
mollification} of the Langevin-Stratonovich equation (\ref{eq8_1_9}).
Case (B): $\lambda(x,t)$ depends explicitly on $x$. Also
in this case 
\begin{equation}
D_S(x,t)= D_{\rm eff}(x,t)
\label{eq8_1_12}
\end{equation}
but the equivalence between the convective contributions
provides the relation
\begin{equation}
v_S(x,t)= v(x)- \frac{D_S(x,t)}{2} \, \partial_x \log \lambda(x,t)
\label{eq8_1_13}
\end{equation}
A physical justification of this phenomenon is addressed
at the end of this paragraph.

The generalization to nonlinear GPK processes in ${\mathbb R}^n$
is straightforward. Define a $N$-state finite Poisson field
$\chi_N({\bf x},t)$
a stochastic process parametrized with respect to ${\bf x} \in {\mathbb R}^n$,
attaining $N$ different possible states, such that the transition
structure between the states is described by the time-continuous
Markov chain
\begin{equation}
\partial_t P_\alpha^{(\chi)}({\bf x},t) = - \lambda_\alpha({\bf x},t)
\, P_\alpha^{(\chi)}({\bf x},t) + \sum_{\gamma=1}^N K_{\alpha,\gamma}({\bf x},t) \, P_\gamma^{(\chi)}({\bf x},t) 
\label{eq8_1_14}
\end{equation}
$\alpha=1,\dots,N$, where  $P_\alpha^{(\chi)}({\bf x},t)$ is the probability
of the occurrence of $\chi_N({\bf x},t)=\alpha$ at position ${\bf x}$
and time $t$.
In equation (\ref{eq8_1_14}) $\lambda_\alpha({\bf x},t)=\sum_{\gamma=1}^N
K_{\gamma,\alpha}({\bf x},t) >0$, and $K_{\alpha,\gamma}({\bf x},t)$
are the entries of a symmetric transition matrix, 
$K_{\alpha,\gamma}({\bf x},t)=K_{\gamma,\alpha}({\bf x},t)$,
for any ${\bf x} \in {\mathbb R}^n$, $t \geq 0$.
Moreover, let us assume   that
\begin{equation}
A_{\alpha,\gamma}({\bf x},t) = \frac{K_{\alpha,\gamma}({\bf x},t)}{\lambda_\gamma({\bf x},t)} 
\label{eq8_1_15}
\end{equation}
represents an irreducible left-stochastic matrix function for any ${\bf x} \in {\mathbb R}^n$ and $t \geq 0$.
Given $N$ vector valued functions ${\bf b}_1({\bf x}), \dots,
{\bf b}_N({\bf x}) : {\mathbb R}^n \rightarrow {\mathbb R}^n$,
satisfying the zero-bias condition
\begin{equation}
\sum_{\alpha=1}^N {\bf b}_\alpha({\bf x})= {\boldsymbol 0}
\label{eq8_1_15bis}
\end{equation}
identically for ${\bf x} \in {\mathbb R}^n$,
a nonlinear GPK process is described by the stochastic differential equation
\begin{equation}
d {\bf x}(t) = {\bf v}({\bf x}(t)) \, dt + {\bf b}_{\chi_N({\bf x},t)}({\bf x})
\, dt
\label{eq8_1_16}
\end{equation}
Its statistical characterization involves $N$ partial
probability density functions $p_\alpha({\bf x},t)$, $\alpha=1,\dots,N$
satisfying the balance equations
\begin{eqnarray}
\partial_t p_\alpha({\bf x},t) & = & - \nabla \cdot \left ( {\bf v}({\bf x}) \,
p_\alpha({\bf x},t) \right )
- \nabla \cdot \left ( {\bf b}_\alpha({\bf x}) \,
p_\alpha({\bf x},t) \right ) \nonumber \\
& - & \lambda_\alpha({\bf x},t) \, p_\alpha({\bf x},t)
+ \sum_{\gamma=1}^N K_{\alpha,\gamma}({\bf x},t) \, p_\gamma({\bf x},t)
\label{eq8_1_17}
\end{eqnarray}

Using the representation in terms of ${\boldsymbol \Lambda}({\bf x},t)=\left (
\lambda_1({\bf x},t),\dots,\lambda_N({\bf x},t) \right )$ and ${\bf A}({\bf x},t)= \left ( A_{\alpha,\beta}({\bf x},t) \right )_{\alpha,\beta=1}^N$, it is  straightforward
to construct a stochastic simulator of eq. (\ref{eq8_1_17})
analogous to that defined in Section 4 of part I  for
linear GPK processes.

It is worth observing  that there is a substantial difference between 
nonlinear Poisson-Kac/GPK process of the form (\ref{eq8_1_1}) or 
 (\ref{eq8_1_16})
and  the nonlinear Langevin equations driven by Wiener perturbations, such
as eq. (\ref{eq8_1_9}).
In the latter case, the ${\bf x}$-dynamics does not
influences the statistical properties
of the stochastic Wiener forcing and implies solely
a modulation of the intensity of the stochastic
perturbation, that depends on ${\bf x}$, via
the factor $\sqrt{2 \, D_S({\bf x},t)}$ as in eq. (\ref{eq8_1_9}).
Conversely, in  the case of Poisson-Kac/GPK processes,
there is a two-way coupling between the ${\bf x}$-dynamics of
the Poissonian perturbation, as the evolution of ${\bf x}(t)$
influences the statistics of the Poissonian field,
whenever the transition rate $\lambda(x,t)$ or
the transition rate vector ${\boldsymbol \Lambda}({\bf x},t)$
depend explicitly on $x$ and ${\bf x}$, respectively.

This observation, physically explain the apparently
``anomalous'' correspondence relation (\ref{eq8_1_13}),
as this model does not fall within the range of application of the
 Wong-Zakai theorem.

\subsection{Continuous GPK processes}
\label{sec_2_2}

A further generalization of GPK theory is the
extension to a continuous number of states. Such a continuous
extension is not suitable within the framework of multi-dichotomic
processes discussed in Section  3 of part I, and this
constitutes the main shortcoming  of this class
of models in the applications to statistical physical problems.

Next, consider the one-dimensional case.
Let $\Xi(t)$ be a time-continuous Markov process attaining
a continuum of states belonging to a domain ${\mathcal D} \in {\mathbb R}$.
Its statistical description involves the transition rate kernel
$K(\alpha,\beta)$, which is a positive symmetric kernel
\begin{equation}
K(\alpha,\beta)=K(\beta,\alpha) \, , \qquad K(\alpha,\beta) \geq 0
\label{eq8_2_1}
\end{equation}
Given $K(\alpha,\beta)$, it is possible to introduce
the transition rates $\lambda(\alpha)$
\begin{equation}
\lambda(\alpha) = \int_{\mathcal D} K(\beta,\alpha) \, d \beta
>0
\label{eq8_2_2}
\end{equation}
and the transition  probability kernel $A(\alpha,\beta)$
\begin{equation}
A(\alpha,\beta)=\frac{K(\alpha,\beta)}{\lambda(\beta)}
\label{eq8_2_3}
\end{equation}
The transition  probability kernel $A(\alpha,\beta)$ possesses the following
properties: (i) normalization, i.e.,
\begin{equation}
\int_{\mathcal D} A(\beta,\alpha) \, d \beta=1
\label{eq8_2_3bis}
\end{equation}
i.e., it is a left-stochastic kernel, and (ii) it is assumed
that $A(\alpha,\beta)$ is irreducible,
meaning  that  solely the constant function $1_{\mathcal D} \in {\mathcal D}$ is the left eigenfunction of $A(\alpha,\beta)$, associated
with the Frobenius eigenvalue $1$. In other terms, the multiplicity
of the Frobenius eigenvalue is $1$.

Indicating with $P^{(\Xi)}(\alpha,t) \, d \alpha = \mbox{Prob} \left [
\Xi(t) \in (\alpha,\alpha+d \alpha) \right ]$, 
$\int_{\mathcal D} P^{(\Xi)}(\alpha,t) \, d \alpha=1$ for any $t \geq 0$,
the evolution of this probability density follows the Markovian
character of the transition dynamics
\begin{equation}
\partial_t p^{(\Xi)}(\alpha,t) = -
\left ( \int_{\mathcal D} K(\beta,\alpha) \, d \beta \right )
\, p^{(\Xi)}(\alpha,t) + \int_{\mathcal D} K(\alpha,\beta)
\, p^{(\Xi)}(\beta,t) \, d \beta
\label{eq8_2_4}
\end{equation} 

Let $b(x,\alpha) : {\mathbb R} \times {\mathcal D} \rightarrow {\mathbb R}$,
the stochastic velocity function satisfying the zero-bias
property
\begin{equation}
\int_{\mathcal D} b(x,\alpha) \, d \alpha =0
\label{eq8_2_4bis}
\end{equation}
for any $x \in {\mathbb R}$. 
Within this setting, it is possible to introduce the
stochastic differential equation
\begin{equation}
d x(t)= v(x) \, d t + b(x,\Xi(t)) \, dt
\label{eq8_2_5}
\end{equation}
which represents the microdynamic description
of a one-dimensional Continuous GPK process (CGPK).
The statistical description of a CGPK process involves the partial
probability density functions $p(x,t;\alpha)$ continuously
parametrized with respect to the  state variable $\alpha$
 of the stochastic forcing $\Xi(t)$,
\begin{equation}
p(x,t;\alpha) \, d x \, d \alpha = \mbox{Prob} \left [
X(t) \in (x,x+dx) \, , \Xi(t) \in (\alpha,\alpha+d \alpha) \right ]
\label{eq8_2_6}
\end{equation}
such that the overall probability density function
for $X(t)$ and its diffusive flux are respectively given by
\begin{equation}
p(x,t)= \int_{\mathcal D} p(x,t;\alpha) \, d \alpha
\, , \qquad
J_d(x,t) = \int_{\mathcal D} b(x,\alpha) \, p(x,t;\alpha) \, d \alpha
\label{eq8_2_7}
\end{equation}
In the continuous setting
$p(x,t;\alpha)$ represents the primitive statistical description of
 a CGPK process, and its evolution equation is expressed by
\begin{eqnarray}
\partial_t p(x,t;\alpha) & = & - \partial_x \left ( v(x) \, p(x,t;\alpha)
\right ) - \partial_x \left   ( b(x,\alpha) \, p(x,t,;\alpha) \right )
\nonumber \\
& + & \int_{\mathcal D} K(\alpha,\beta) \, \left [ p(x,t;\beta)
- p(x,t;\alpha) \right ] \, d \beta
\label{eq8_2_8}
\end{eqnarray}
where the symmetry of the kernel $K(\alpha,\beta)$ has been
applied.

Assume for simplicity that $b(x,\alpha)$ does not depend on $x$,
and that the stochastic velocity function $b(\alpha)$ is a monotonic
function of $\alpha$.  Under these conditions one can simply
map the states of $\Xi(t)$ using the transformation
$c=b(\alpha)$, thus defining the new stochastic process $\Theta(t)$
defined as
\begin{equation}
b(\Xi(t)) = \Theta(t) \, , \qquad \Xi(t)=b^{-1}(\Theta(t))
\label{eq8_2_9}
\end{equation}
Let ${\mathcal C}=b({\mathcal D})$.
In this way,  the stochastic differential equation
defining CGPK process attains
the more compact expression
\begin{equation}
d x(t)= v(x(t)) \, d t + \Theta(t) \, dt
\label{eq8_2_10}
\end{equation}
where $\Theta(t) \in {\mathcal C}$.
Letting $f(c)=b^{-1}(c)$ the inverse of $b(\alpha)$, and define
\begin{equation}
\overline{p}(x,t;c)= p(x,t;\alpha) |_{\alpha=f(c)} \, \frac{d f(c)}{d c}
\label{eq8_2_10bis}
\end{equation}
\begin{equation}
\overline{K}(c,c^\prime) = K(\alpha,\beta) |_{\alpha=f(c), \,
\beta=f(c^\prime)} \, 
\frac{d f(c)}{d c}
\label{eq8_2_10tris}
\end{equation}
the CGKP process,  parametrized with respect to the values $c$
attained by stochastic velocity  $b(\alpha)$, 
is described by the transformed probability
density functions $\overline{p}(x,t;c)$ associated
with eq. (\ref{eq8_2_10}), that satisfy the balance
equations
\begin{eqnarray}
\partial_t \overline{p}(x,t;c) & = & - \partial_x \left (
v(x) \, \overline{p}(x,t;c)  \right ) - c \, \partial_x \overline{p}(x,t;c)
\nonumber \\
& + & \left ( \int_{\mathcal C} \overline{K}(c^\prime,c) \, d c^\prime
\right ) \, \overline{p}(x,t;c) - \int_{\mathcal C} \overline{K}(c,c^\prime) \, \overline{p}(x,t;c^\prime) \, d c^\prime
\label{eq8_2_11}
\end{eqnarray}
The transformation of the transition kernel
(\ref{eq8_2_10tris}) is such that the transition
rates $\overline{\lambda}(c)$ coincide with
the corresponding ones expressed with respect to the 
parameter $\alpha$, namely
\begin{equation}
\overline{\lambda}(c) = \int_{\mathcal C} {\overline K}(c^\prime,c)
\, d c^\prime = \int_{\mathcal D} K(\beta, f(c)) \, d \beta=
\lambda(\alpha) |_{\alpha=f(c)}
\label{eq8_2_11bis}
\end{equation}
while the transformed transition probability matrix kernel
$\overline{A}(c^\prime,c)$ is defined as
\begin{equation}
\overline{A}(c^\prime,c) = \frac{\overline{K}(c^\prime,c)}{\overline{\lambda}(c)} = A(\beta,\alpha) |_{\alpha=f(c), \, \beta=f(c^\prime)} \, \frac{d f(c^\prime)}{d c^\prime}
\label{eq8_2_11tris}
\end{equation}
Observe that if $K(\alpha,\beta)$ is symmetric, this is no
longer true for $\overline{K}(c,c^\prime)$ in the presence
of a nonlinear expression of $b(\alpha)$.

In the case $b(\alpha)=\alpha$, and $K(\alpha,\beta)$ symmetric,
the analysis greatly simplifies, and  eq.  (\ref{eq8_2_11}) reduces
to
\begin{eqnarray}
\partial_t \overline{p}(x,t;c) & = & - \partial_x \left (
v(x) \, \overline{p}(x,t;c)  \right ) - c \, \partial_x \overline{p}(x,t;c)
\nonumber \\
& + &   \int_{\mathcal D} \overline{K}(c,c^\prime) 
\left [  \overline{p}(x,t;c^\prime) -  \overline{p}(x,t;c) \right ]
\, d c^\prime
\label{eq8_2_11z}
\end{eqnarray}
since ${\mathcal C}={\mathcal D}$.
It can be recognized that eq. (\ref{eq8_2_11z})
represents a one-dimensional linearized Boltzmann equation.
The zero-bias condition (\ref{eq8_2_4bis}) becomes 
\begin{equation}
\int_{\mathcal B} c \, d c =0
\label{eq8_2_12}
\end{equation}
which indicates that the domain ${\mathcal C}$ must be the
union of  symmetric intervals with respect to $c=0$, eventually
reduces to a single interval
${\mathcal C}={\mathcal B}=(-b^{(c)},b^{(c)})$. 
The overall probability density function $\overline{p}(x,t)=\int_{\mathcal B}
\overline{p}(x,t;c) \, d c$
satisfies the balance equation
\begin{equation}
\partial_t \overline{p}(x,t) = - \partial_x \left ( v(x)  \, \overline{p}(x,t;c)
\right ) - \partial_x \overline{J}_d(x,t)
\label{eq8_2_13}
\end{equation}
where the diffusive flux $\overline{J}_d(x,t)$ fulfills the
constitutive equation
\begin{eqnarray}
\partial_t \overline{J}_d(x,t) = - \partial_x \left ( v(x) \, \overline{J}_d(x,t) \right ) - \partial_x \int_{\mathcal B} c^2 \, \overline{p}(x,t;c) \, d c
\nonumber \\
-\int_{\mathcal B} c \, \overline{\lambda}(c) \, \overline{p}(x,t;c) \, d c
+ \int_{\mathcal B} \int_{\mathcal B} c \, \overline{K}(c, c^\prime)
\, \overline{p}(x,t,c^\prime) \, d c \, d c^\prime
\label{eq8_2_14}
\end{eqnarray}
where $\overline{\lambda}(c)=\int_{\mathcal B} K(c^\prime, c) \, d c$.

The Kac limit can be ascertained  in the continuous case
using techniques and arguments analogous to those developed in part I.
 To give an example, assume ${\mathcal B}=(-b^{(c)},b^{(c)})$,
$\overline{\lambda}(x)=\overline{\lambda}=\mbox{constant}$,
and that there exits a $\delta <1$ such that
\begin{equation}
\int_{\mathcal B} c \, \overline{K}(c,c^\prime) \, d c= \delta \, c^\prime
\label{eq8_2_15}
\end{equation}
For $b^{(c)}, \, \overline{\lambda} \rightarrow \infty$, keeping fixed
the nominal diffusivity $D_{\rm nom}=(b^{(c)})^2/2 \overline{\lambda}$, 
the infinitely
fast recombination  implies the equipartition amongst the
partial  probability waves,
\begin{equation}
\overline{p}(x,t;c) = \frac{\overline{p}(x,t)}{2 \, b^{(c)}} + o\left (
\overline{\lambda}^{-1} \right )
\label{eq8_2_16}
\end{equation}
and the effective diffusivity can be obtained in closed form
\begin{equation}
D_{\rm eff}= \frac{1}{2 \, (1-\delta) \, \overline{\lambda} \, b^{(c)}}
\, \int_{\mathcal B} c^2 \, d c= \frac{2 \, D_{\rm nom}}{3 \, (1- \delta)}
\label{eq8_2_17}
\end{equation}
Other cases of interest  are addressed in paragraph \ref{sec_2_4}.

\subsection{Nonlinear FPK models}
\label{sec_2_3}

In paragraph \ref{sec_2_1} we have indicated with the diction
{\em nonlinear GPK} those processes in which, either the stochastic
velocity vectors, or the transition rates/transition probability
matrix, or both, depend on the
state variable ${\bf x}$.
There is another, significant, source of nonlinearity
in stochastic models occurring whenever the statistical
properties of the stochastic perturbation influence the dynamics
of the stochastic process ${\bf X}(t)$ itself, so that
the stochastic microdynamics depends on the collective
behavior of ${\bf X}(t)$, i.e., on its probability density functions.
This is the case of the nonlinear Fokker-Planck equation
introduced by McKean for Langevin-Wiener stochastic models \cite{mckean}.
For review see \cite{nonlinearfokker}.
In the case of Poisson-Kac and GPK model this means
that the stochastic velocity vectors and the transition rates
are functionals of the probability densities describing ${\bf X}(t)$,
i.e., of the partial probability density functions $\{p_\alpha({\bf x},t)\}_{\alpha=1}^N$.
We refer to this situation as a {\em Nonlinear Fokker-Planck-Kac}
model (NFPK).

The paradigm of  Nonlinear FPK processes is represented by the one-dimensional Poisson-Kac model
\begin{equation}
d x (t) = v(x(t)) \, d t + b (-1)^{\chi(x,t,\{p^\pm(x(t),t)\})} \, dt
\label{eq8_3_1}
\end{equation}
where $\chi(x,t;\{p^\pm(x(t),t)\})$ is a Poisson field the transition
rate of which $\lambda(x,\{p^\pm(x(t),t)\})$ is a positive
functional of the partial probability waves associated with
${\bf X}(t)$. For example, it can depend linearly on
$p^{\pm}(x,t)$, such as
\begin{equation}
\hspace{-1.0cm} 
\lambda(x,\{p^\pm(x(t),t)\})= \lambda_0(x) + \int_{\mathbb R} a^+(x) \,
p^+(x,t) \, d x + \int_{\mathbb R} a^-(x) \, p^-(x,t) \, d x
\label{eq8_3_2}
\end{equation}
where $\lambda_0(x), \, a^+(x),\, a^-(x)$ are functions of $x$,
or can be a function of the partial moment hierarchy associated with 
$p^\pm(x,t)$.
For the NFPK process (\ref{eq8_3_1}), the evolution
equations for the partial probability  waves become nonlinear, namely
\begin{equation}
\hspace{-1.5cm} 
\partial_t p^\pm(x,t)  =  - \partial_x \left [(v(x) \pm b) \, p^\pm(x,t)
\right ]
 \mp  \lambda(x,\{p^\pm(x(t),t)\}) \left [ p^+(x,t)-p^-(x,t) \right ]
\label{eq8_3_3}
\end{equation}
Alternatively, another class of NGPK processes can be defined
in the case the advective contribution $v(x,\{p^\pm(x(t),t)\})$ represents a functional of the partial probability waves.
An example of this class of model is addressed in the
next paragraph.
 
The generalization to GPK processes in ${\mathbb R}^n$ is straightforward.
We develop an example in the next Section, combining
all the extension discussed so far in order to
address a physically relevant problem, namely the
stochastic nature of the collisional Boltzmann equation.

\subsection{Examples}
\label{sec_2_4}

In this paragraph we analyze three prototypical examples
covering the range of generalization of GPK processes
treated in the previous paragraphs.

\begin{figure}[h]
\begin{center}
{\includegraphics[height=6cm]{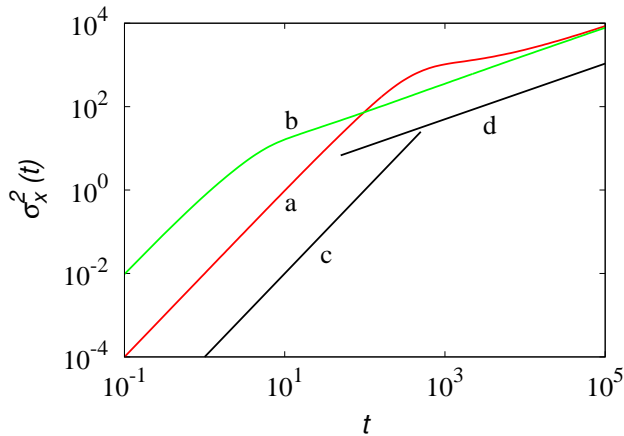}}
\end{center}
\caption{$\sigma_x^2(t)$ vs $t$ for the nonlinear Poisson-Kac process
associated with the state-dependent transition rate (\ref{eq8_4_2}) 
at $D_{\rm nom}=1$.
Line (a) refers to $b=0.1$, line (b) to $b=1$.
Lines (c) and (d) represent the scalings $\sigma_x^2(t) \sim t^2$,
and $\sigma_x^2(t) \sim t^{2/3}$, respectively.}
\label{Fig16}
\end{figure}

To begin with, consider a one-dimensional
Poisson-Kac process in ${\mathbb R}$, defined by
the stochastic differential equation
\begin{equation}
d x(t)= b \, (-1)^{\chi(x,t)} \, dt
\label{eq8_4_1}
\end{equation}
where the Poisson field $\chi(x,t)$ is characterized by
the following transition-rate function
\begin{equation}
\lambda(x)= \lambda_0 (1+ |x| )
\label{eq8_4_2}
\end{equation}
Figure \ref{Fig16} depicts the behavior of the mean square
displacement $\sigma_x^2(t)$ vs $t$, obtained from the stochastic
simulation of eqs. (\ref{eq8_4_1})-(\ref{eq8_4_2}),
using $N_p=10^6$ particles starting from $x=0$ at 
$D_{\rm nom}=b^2/2 \lambda_0=1$,
 for two different values of $b$: $b=0.1$ (line a) and $b=1$ (line b).
In this problem $\langle x(t) \rangle =0$, so that $\sigma_x^2(t)=
\langle x^2(t) \rangle$.

The mean square displacement $\sigma_x^2(t)$ admits a crossover
behavior, 
\begin{equation}
\sigma_x^2(t) \sim
\left \{
\begin{array}{lll}
t^2 & & t < t^* \\
t^{2/3} &  & t \gg t^*
\end{array}
\right .
\label{eq8_4_3}
\end{equation}
where the crossover time $t^*$ is about $10^4$ at $b=0.1$, and order
of $10^1$ at $b=1$. At short time-scales $\sigma_x^2(t)$
grows quadratically with time $t$, due to
the finite propagation velocity (a.e. smoothness)
 characterizing Poisson-Kac and GPK
processes. Asymptotically, i.e., for $t \gg t^*$, the 
 mean square displacement exhibits an anomalous scaling
in time with an exponent equal to $2/3$.

This result can be easily interpreted using elementary
scaling analysis, as the transition
rate expression (\ref{eq8_4_2}) corresponds to
a position dependent diffusion coefficient $D_{\rm eff} \sim 1/|x|$.
Consequently, $\sigma_x^2(x) \sim L^2 \sim D_{\rm eff}(L) \, t \sim t \, L$,
and therefore $L \sim t^{1/3}$, implying $L^2 \sim \sigma_x^2(t) \sim t^{2/3}$.

This example admits another byproduct: in the framework
of nonlinear GPK processes it is possible to generate anomalous diffusion
scalings just by a suitable choice of the transition rate function $\lambda(x)$
characterizing the  Poissonian field $\chi(x,t)$
or $\chi_N({\bf x},t)$.

\begin{figure}[h]
\begin{center}
{\includegraphics[height=8cm]{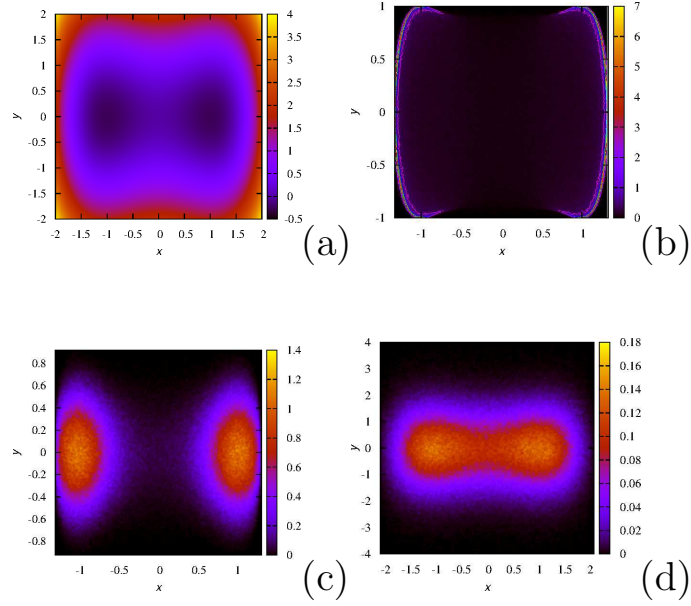}}
\end{center}
\caption{Panel (a): contour plot of the potential $U(x,y)$ eq. (\ref{eq8_4_4}).
Panels (b) to (d): contour plots of the stationary densities
$p^*(x,y)$ for the Continuous GPK process (\ref{eq8_4_6})
at different values of the parameters. Panel
(b): $D_{\rm eff}=1$, $b=1$, panel (c): $D_{\rm eff}=0.1$,
$b=1$, panel (d): $D_{\rm eff}=1$, $b=10$.}
\label{Fig17}
\end{figure}

As a second example, consider a continuous GPK process in
${\mathbb R}^2$ corresponding to the dynamics, in the overdamped
regime, of a particle moving in the  potential
\begin{equation}
U({\bf x})=U(x,y)= \frac{x^4}{4} -\frac{x^2}{2} + \frac{y^2}{2}
= U_x(x)+U_y(y)
\label{eq8_4_4}
\end{equation}
the contour plot of which is depicted in  figure \ref{Fig17} panel (a).
This expression corresponds to the superposition of a bistable potential
$U_x(x)$ along the $x$-coordinate, and of a harmonic potential in the 
$y$-coordinate. Assume a friction factor $\eta=1$ a.u., and let
$\Xi(t)$ a continuous Markov process attaining values in $[0,2 \pi)$,
characterized by a uniform transition rate $\lambda$ and by
a uniform transition kernel
\begin{equation}
K(\alpha,\beta) = \frac{\lambda}{2 \, \pi} \qquad \alpha,\beta \in [0, 2 \pi)
\label{eq8_4_5}
\end{equation}
Consider the Continuous GPK process defined
by the stochastic differential equation
\begin{equation}
d {\bf x}(t) = - \nabla_x U({\bf x}(t)) \, dt + {\bf b}(\Xi(t)) \, dt
\label{eq8_4_6}
\end{equation}
where the stochastic velocity function ${\bf b}(\alpha)$ is defined
by
\begin{equation}
{\bf b}(\alpha) = b \, \left ( \cos(\alpha), \sin(\alpha) \right ) \qquad \alpha \in [0,2 \pi)
\label{eq8_4_7}
\end{equation}
For this problem, the balance equation for the partial probability
waves $p({\bf x},t;\alpha)$ reads
\begin{eqnarray}
\partial_t p({\bf x},t;\alpha) & = & \nabla_x  \cdot \left (
\nabla_x U({\bf x}) \, p({\bf x},t;\alpha)  \right ) - {\bf b}(\alpha) \cdot
\nabla_x p({\bf x},t;\alpha) \nonumber \\
& - & \lambda \, p({\bf x},t;\alpha) + \frac{\lambda}{2 \, \pi} \int_0^{2 \pi}
p({\bf x},t;\beta) \, d \beta
\label{eq8_4_8}
\end{eqnarray}
In this case, the diffusive flux is given by
${\bf J}_d({\bf x},t)= \int_0^{2 \pi}   {\bf b}(\alpha) \, p({\bf x},t;
\alpha) \, d \alpha$. For  the dyadic tensor associated with ${\bf b}(\alpha)$
one has
\begin{equation}
\int_0^{2 \pi} {\bf b}(\alpha) \, {\bf b}(\alpha) \, d \alpha = 
\frac { b^2 \, {\bf I}}{2} 
\label{eq8_4_9}
\end{equation}
where ${\bf I}$ is the identity matrix, and consequently
 the Kac limit provides
$D_{\rm eff}=D_{\rm nom}=b^2/2 \lambda$.
The numerical simulation of this stochastic process possessing
a continuum of  states $\alpha$ is simple: (i)  the
intervals between two consecutive transitions $\tau$
are distributed exponentially
 with probability density $p_\tau(\tau)=\lambda
\, e^{-\lambda \, \tau}$, (ii) whenever a transition occurs from a state
$\alpha$ to a new state, say $\beta$, the determination of the new
state $\beta$ is chosen randomly from a uniform probability distribution
in $[0,2 \pi)$.

\begin{figure}[h]
\begin{center}
{\includegraphics[height=10cm]{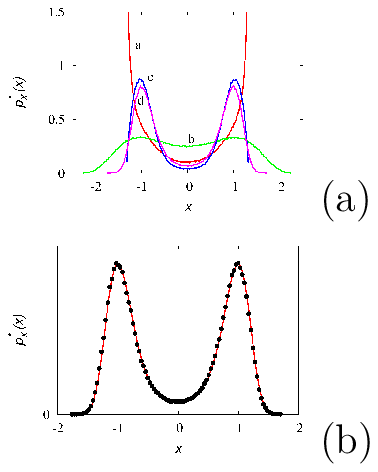}}
\end{center}
\caption{Panel (a): Stationary marginal probability densities $p_x^*(x)$
vs $x$ associated with the Continuous GPK process (\ref{eq8_4_6}).
Line (a) refers to $D_{\rm eff}=1$, $b=1$,
line (b) to $D_{\rm eff}=1$, $b=10$,
line (c) to $D_{\rm eff}=0.1$, $b=1$,
line (d) to $D_{\rm eff}=0.1$, $b=10$.
Panel (b): Comparison of the stationary solution of the continuous
GPK process at $D_{\rm eff}=0.1$, $b=10$, symbols ($\bullet$),
and the corresponding Boltzmannian distribution 
$p^*(x)= \exp(-U_x(x)/D_{\rm eff})$.}
\label{Fig18}
\end{figure}

Figure \ref{Fig17} panels (b)-(d) depict the contour plot of the
stationary overall probability density functions $p^*(x,y)$
for several values of $D_{\rm eff}$ and $b$, obtained from
stochastic simulations of eq. (\ref{eq8_4_6}) using
an ensemble of $10^6$ particles.
The corresponding stationary marginal distributions $p^*_x(x)=
\int_{\mathbb R} p^*(x,y) \, dy$ with respect to the $x$ coordinate
are depicted in figure \ref{Fig18} panel (a).
At low values of $D_{\rm eff}$ and $b$, panel (b) in figure \ref{Fig17}
and curve (a) in figure \ref{Fig18}, the
stationary probability density concentrates
in an external shell far away from the minima of the
potential. This is a peculiar feature of undulatory
transport models in the presence of conservative potentials, 
which may possess an invariant region
$\Omega$ such that $(-\nabla_x U({\bf x}) +{\bf b}(\alpha)) \cdot {\bf n}_e({\bf x}) |_{\partial \Omega} \leq 0$, where  ${\bf n}_e({\bf x})$
is the external normal unit vector at points ${\bf x} \in \partial \Omega$
(see also the discussion in part II on invariant regions).

As $D_{\rm eff}$ decreases, $p^*(x,y)$ displays the typical
bimodal shape characterizing bistable motion with diffusion,
panel (d) in figure \ref{Fig17} and lines (b) and (d) in figure \ref{Fig18}. 
As $b$ increases, keeping constant the value of the effective
diffusivity, the stationary density
approaches the Boltzmannian distribution. An example
is depicted in figure  \ref{Fig17} panel (d) and it is
clearly evident from figure \ref{Fig18} panel (b) that 
the marginal stationary distribution $p_x^*(x)$, 
at $D_{\rm eff}=0.1$ and $b=10$, practically coincides with
its Boltzmann-Kac limit $p_x^*(x)= A \, e^{-U_x(x)/D_{\rm eff}}$
(observe that $\eta=1$).

Finally, consider a Nonlinear FPK process expressed in the form
of a Shimizu-Yamada model \cite{shimizu}. More precisely, consider the
stochastic one-dimensio\-nal differential equation
\begin{equation}
d x(t) = v(x(t),\{p^\pm(x(t),t)\}) \, dt + b (-1)^{\chi(t)} \, dt
\label{eq8_4_10}
\end{equation}
where $\chi(t)$ is a usual Poisson process possessing constant
transition rate $\lambda$, and the drift velocity
depends on the partial probability densities as
\begin{equation}
\hspace{-1.0cm} 
v(x,\{p^\pm(x,t)\}) = - x+ \varepsilon \left [ m_+^{(1)}(t)+m_-^{(1)}(t)
\right ] + v_f = -x + \varepsilon \, m^{(1)}(t) + v_f
\label{eq8_4_11}
\end{equation}
where $m^{(1)}_{\pm}(t)= \int_{\mathbb R} x \, p^{\pm}(x,t) \, dx$
are the partial first-order moments of the process, and $\varepsilon$,
$v_f$ two constant parameters.
This model corresponds to the dynamics of a stochastic particle
in a harmonic potential (in the overdamped regime), subjected to
an external constant bias $v_f$ and to an additional contribution
proportional to the overall first-order moment   $m^{(1)}(t)$.

\begin{figure}[h!]
\begin{center}
{\includegraphics[height=12cm]{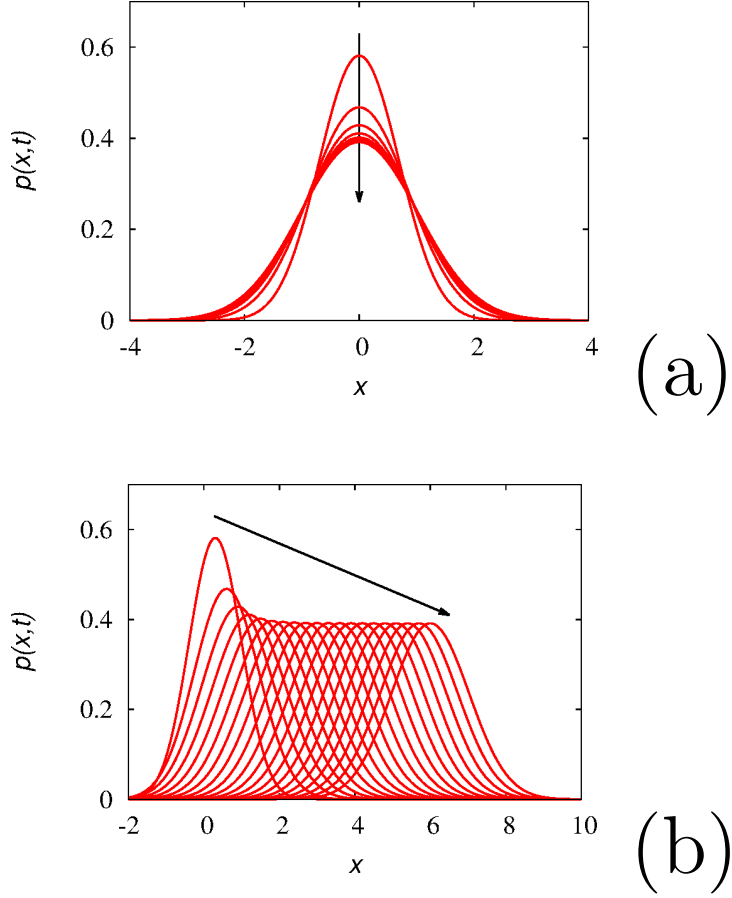}}
\end{center}
\caption{Evolution of $p(x,t)$ vs $t$ for
the nonlinear FPK process (\ref{eq8_4_10})-(\ref{eq8_4_11})
 at $D_{\rm eff}=1$,
$\lambda=100$ starting from an initially impulsive
condition (\ref{eq8_4_13}), for  two different  sets of  parameter
values 
 entering eq. (\ref{eq8_4_11}).  The profiles are sampled at uniform 
time intervals $t_n=n \Delta t$, $n=1,..21$, $\Delta t=0.3$.
 Panel (a) refers to 
$\varepsilon=v_f=0$;
 panel (b) to $\varepsilon=v_f=1$.}
\label{Fig19}
\end{figure}

The balance equation for the partial waves attains the form
\begin{equation}
\hspace{-1.0cm}
\partial_t p^{\pm}(x,t) = \partial_x \left [(x-\varepsilon \, m^{(1)}(t)
-v_f \mp b ) \, p^{\pm}(x,t) \right ]
\mp \lambda \left [p^+(x,t)-p^-(x,t) \right ]
\label{eq8_4_12}
\end{equation}
where $m^{(1)}(t)=m_+^{(1)}(t)+m_-^{(1)}(t)$. 
As initial condition for the partial probability waves consider
\begin{equation}
p^{\pm}(x,0)= \frac{\delta(x)}{2}
\label{eq8_4_13}
\end{equation}
Figure \ref{Fig19}  panel (a) depicts the evolution
of the overall density function $p(x,t)$
in the case of  a pure harmonic oscillator $\varepsilon=v_f=0$
(linear case)
at $D_{\rm eff}=1$, $\lambda=100$, for which the density function
becomes localized at $x=0$ with a variance equal to $D_{\rm eff}$.
Panel (b) refers to $\varepsilon=1$, $v_f=1$ i.e., to a
truly Nonlinear FPK model.

\begin{figure}[h!]
\begin{center}
{\includegraphics[height=5cm]{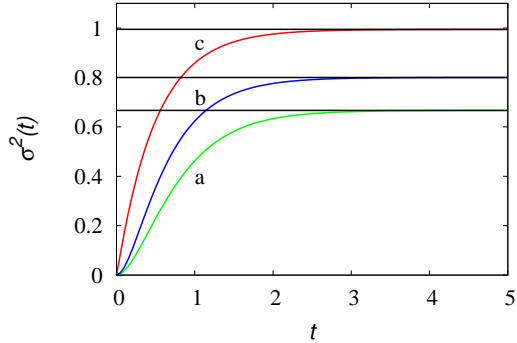}}
\end{center}
\caption{$\sigma^2(t)$ vs $t$ for the nonlinear FPK model
discussed in the main text at $D_{\rm eff}=1$, for different values
of $\lambda$. 
Line (a) refers to $\lambda=1$, (b) to $\lambda=2$, (c) to
$\lambda=100$. The horizontal lines
correspond to the predictions of eq. (\ref{eq8_4_22}).}
\label{Fig20}
\end{figure}

In order to derive the salient qualitative properties 
of this model consider the first elements of the partial
moment hierarchy.
For the zero-th order moments, the symmetric initial
condition (\ref{eq8_4_13}) implies
\begin{equation}
m_+^{(0)}(t) = m_-^{(0)}(t)= \frac{1}{2} \qquad t \geq 0
\label{eq8_4_14}
\end{equation}
The evolution equations for the first-order moments attain the form
\begin{equation}
\hspace{-1.0cm}
\partial_t m_{\pm}^{(1)}(t) = - m_\pm^{(1)}(t) + \left [
m^{(1)}(t)+ v_f \pm b \right ] \, m_\pm^{(0)}(t) -\lambda
\left [ m_+^{(1)}(t) -m_-^{(1)}(t) \right ]
\label{eq8_4_15}
\end{equation}
Summing the equations for $m_+^{(1)}(t)$ and $m_-^{(1)}(t)$,
and accounting for eq. (\ref{eq8_4_14})
one obtains
\begin{equation}
\partial_t m^{(1)}(t) = v_f
\label{eq8_4_16}
\end{equation}
which implies 
\begin{equation}
m^{(1)}(t) = v_f \, t
\label{eq8_4_17}
\end{equation}
Therefore, the probability  profile moves at constant speed $v_f$.
As regards the partial second-order moments
\begin{equation}
\hspace{-1.5cm}
\partial_t  m_\pm^{(2)}(t) = - 2 \, m_\pm^{(2)}(t)
+  2 \, \left [
m^{(1)}(t)+ v_f \pm b \right ] \, m_\pm^{(1)}(t) -\lambda
\left [ m_+^{(2)}(t) -m_-^{(2)}(t) \right ]
\label{eq8_4_18}
\end{equation}
From eqs. (\ref{eq8_4_14})-(\ref{eq8_4_18}),
the overall variance $\sigma^2(t)=m_+^{(2)}(t)+ m_-^{(2)}(t)
- \left [ m_+^{(1)}(t)+m_-^{(1)}(t) \right ]^2$ fulfills the
balance equation
\begin{equation}
\partial_t \sigma^2(t) = - 2 \, \sigma^2(t) + 2 \, b \left [
m_+^{(1)}(t) - m_-^{(1)}(t) \right ]
\label{eq8_4_19}
\end{equation}
Taking the difference  between the evolution
equations of the first-order
partial moments provides
\begin{equation}
\hspace{-1.5cm}
\partial_t \left [
m_+^{(1)}(t) - m_-^{(1)}(t) \right ]
= - \left [
m_+^{(1)}(t) - m_-^{(1)}(t) \right ]
+ b - 2 \, \lambda \, \left [
m_+^{(1)}(t) - m_-^{(1)}(t) \right ]
\label{eq8_4_20}
\end{equation}
Asymptotically, the difference between the partial first-order
moments converges towards the value
\begin{equation}
m_{+,\infty}^{(1)}- m_{-,\infty}^{(1)} = \frac{b}{2 \, \lambda +1}
\label{eq8_4_21}
\end{equation}
which implies that
 $\lim_{t \rightarrow \infty} \sigma^2(t)=\sigma_\infty^2$
where
\begin{equation}
\sigma_\infty^2 = \frac{b^2}{2 \, \lambda +1}= \frac{D_{\rm eff}}{1+ 1/2 \lambda}
\label{eq8_4_22}
\end{equation}
The variance of the overall probability density
 wave attains asymptotically a constant
value equal to $\sigma_\infty^2$. The NFPK considered
above describes the evolution of a nonlinear soliton
traveling with constant speed $v_f$ and possessing a constant
variance $\sigma_\infty^2$ (as can be observed from
the profiles in figure \ref{Fig19} panel (b)).
Figure \ref{Fig20} depicts the comparison of numerical
simulation results for $\sigma^2(t)$ and the asymptotic
expression (\ref{eq8_4_22}).

\begin{figure}[h!]
\begin{center}
{\includegraphics[height=5cm]{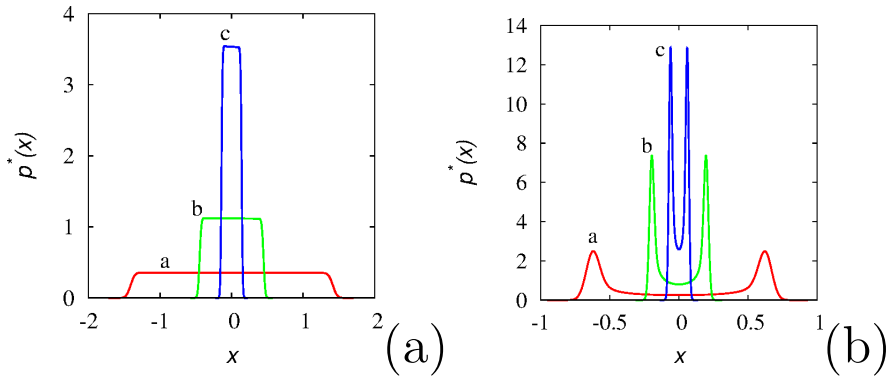}}
\end{center}
\caption{Shape of the propagating soliton pulse. Panel (a)
refers to $\lambda=1$, panel (b) to $\lambda=0.2$.
Lines (a) to (c) refer to $D_{\rm eff}=1, \, 10^{-1},\, 10^{-2}$,
respectively.}
\label{Fig21}
\end{figure}

The shape of the propagating solitons depends significantly
on the transition rate $\lambda$. For high values of $\lambda$,
a nearly Gaussian soliton propagates as expected from the
Kac limit, see figure \ref{Fig19} panel (b). However, for small values
 of $\lambda$,    profiles  completely different
from the Gaussian one 
can occur. This phenomenon is depicted
in figure \ref{Fig21} panels (a) and (b), corresponding to
$\lambda=1$ and $\lambda=0.2$, respectively. The resulting
probability density profiles $p^*(x)$ of the propagating solitons,
 depicted in this
figure, are rescaled to unit zero-th order moment,  i.e. $\int_{\mathbb R}
p^*(x) \, dx=1$, and zero
mean.
As can be observed, a nearly rectangular-shaped soliton
occurs for $\lambda=1$ (panel (a)), while bimodal
profiles characterize its shape for lower values of $\lambda$
(panel (b)).

\section{The Boltzmann equation}
\label{sec_3}

Gathering together the generalizations of GPK process
introduced in the previous Section (nonlinearity and continuity
of stochastic states), we arrive at a remarkable
result. To get it, we need two ingredients: (i) a Continuous
GPK process parametrized with respect to the stochastic
velocity vector ${\bf b} \in {\mathcal D} \subseteq {\mathbb R}^n$,
and described by means of the  transition kernel $K({\bf b},{\bf b}^\prime)$. Letting $p({\bf x},t;{\bf b})$
be  the associated partial 
probability densities parametrized with respect to ${\bf b}$, and
assuming for simplicity ${\bf v}({\bf x})=0$,  their
evolution equation is given by
\begin{eqnarray}
\hspace{-1.5cm}
\partial_t p({\bf x},t;{\bf b}) & = & - {\bf b} \cdot
\nabla_x p({\bf x},t;{\bf b})- \int_{\mathcal D}
K({\bf b}^\prime,{\bf b}) \,  d {\bf b}^\prime
 \, p({\bf x},t;{\bf b})  
 \nonumber \\
& + & \int_{\mathcal D}
K({\bf b},{\bf b}^\prime) \, p({\bf x},t;{\bf b}^\prime)  d {\bf b}^\prime
\label{eq_8x_2}
\end{eqnarray}
 (ii) the further assuption that this Continuous GPK
process is a Nonlinear FPK process, in which the
transition kernel $K({\bf b},{\bf b}^\prime;[p])$ is
a linear homogeneous functional of the partial
probability waves $p({\bf x},t;{\bf b})$, i.e.,
\begin{equation}
K({\bf b},{\bf b}^\prime; [p])= \int_{{\mathcal D}} \int_{{\mathcal D}}
H({\bf b}, {\bf b}^{\prime \prime \prime} | {\bf b}^\prime, {\bf b}^{\prime \prime})
\, p({\bf x},t;{\bf b}^{\prime \prime}) \, d {\bf b}^{\prime \prime}
\,  d {\bf b}^{\prime \prime \prime}
\label{eq_8b_1}
\end{equation}
where the kernel $H({\bf b}, {\bf b}^{\prime \prime \prime} | {\bf b}^\prime, {\bf b}^{\prime \prime})$ fulfills the symmetry
\begin{equation}
H({\bf b},{\bf b}^{\prime \prime \prime}  | {\bf b}^\prime,{\bf b}^{\prime \prime}) =
H({\bf b}^\prime,{\bf b}^{\prime \prime} | {\bf b}, {\bf b}^{\prime \prime \prime})
\label{eq_8x_3}
\end{equation}
In the absence of a deterministic biasing field,
the balance equation for the partial probability density
$p({\bf x},t;{\bf b})$ associated with  the above-defined
Nonlinear CGPK process reads
\begin{equation}
\hspace{-2.3cm}
\partial_t p({\bf x},t;{\bf b}) = - {\bf b} \cdot 
\nabla_x p({\bf x},t;{\bf b}) + \int_{{\mathcal D}}
\int_{{\mathcal D}} \int_{{\mathcal D}}
 H({\bf b},{\bf b}^{\prime \prime \prime} | {\bf b}^\prime,{\bf b}^{\prime \prime})
\, \left [ p^\prime \, p^{\prime \prime} - p  \,p^{\prime \prime \prime} \right ]
d {\bf b}^\prime \, d {\bf b}^{\prime \prime} \,  d {\bf b}^{\prime \prime \prime}
\label{eq_8b_2}
\end{equation}
where $p=p({\bf x},t;{\bf b})$, $p^\prime=p({\bf x},t;{\bf b}^\prime)$
and $p^{\prime \prime}=p({\bf x},t;{\bf b}^{\prime \prime})$,
$p^{\prime \prime \prime}=p({\bf x},t;{\bf b}^{\prime \prime \prime})$.

Eq. (\ref{eq_8b_2}) corresponds  to the nonlinear Boltzmann equation
once the expression for the ``collision''
 kernel $H({\bf b},{\bf b}^{\prime \prime \prime} | {\bf b}^\prime 
{\bf b}^{\prime \prime})$ 
has been  defined consistently with the conservation laws (momentum and
kinetic energy) associated with the elastic nature of the collisions 
\cite{balescu}, i.e.,
\begin{equation}
\hspace{-2.3cm}
H({\bf b},{\bf b}^{\prime \prime \prime} | {\bf b}^\prime,{\bf b}^{\prime \prime})
= G({\bf b},{\bf b}^{\prime \prime \prime} | {\bf b}^\prime,{\bf b}^{\prime \prime})
\, \delta({\bf b}+ {\bf b}^{\prime \prime \prime}- {\bf b}^\prime-{\bf b}^{\prime \prime}) \, \delta(|{\bf b}|^2+ |{\bf b}^{\prime \prime \prime}|^2- |{\bf b}^\prime|^2-|{\bf b}^{\prime \prime}|^2)
\label{eq_8b_agg2}
\end{equation}
where the kernel $G({\bf b},{\bf b}^{\prime \prime \prime} | {\bf b}^\prime,{\bf b}^{\prime \prime})$ accounts for the particle scattering cross section
and possesses obvious symmetries.

In the GPK framework, the partial probability density waves
$p({\bf x},t;{\bf b})$ correspond to the one-particle distribution function
$f({\bf x},{\bf v},t)$ \cite{balescu}, once the particle velocity ${\bf v}$
is identified with the stochastic velocity vector ${\bf b}$
parametrizing the Continuous GPK process.

In eqs. (\ref{eq_8b_1})-(\ref{eq_8b_2}) we assumed
that ${\bf b}$ is defined in the domain ${\mathcal D}$,
which can be either  bounded, in order
to account for the relativistic constraint imposed by light velocity $c$
{\em in vacuo}, and in this case ${\mathcal D}=\{ {\bf b} \; | \; |{\bf b} | \leq c \}$, or ${\mathcal D}={\mathbb R}^n$ as in the
classical Boltzmann equation, where no constraints
are posed on the maximum attainable velocity.
In the CGPK model for the Boltzmann equation, the binary
collisions amongst particles correspond to the
ordinary recombination process amongst the partial  probability waves.

Therefore, the theory of CGPK processes, including nonlinearity
effects in the
transition kernel, provides
a simple and intuitive fully stochastic route to the basic
equation of kinetic theory.
In some sense this result is surprising for its simplicity,
and supports what is usually referred to as the
Kac's program in kinetic theory originated by the 1954-paper
by Kac on the stochastic foundations  of the kinetic theory \cite{kac_kinetic},
in which   a simple one-dimensional
Markovian toy-model  is introduced to describe the relaxation properties
of a rarefied particle gas.
For further details on the Kac's program, and on related recent
advances, the reader is referred to \cite{kacprog1,kacprog2,kacprog3,kacprog4}.

While a thorough analysis of the implications of this result will be 
addressed in future communications, it is important to
point out several observations of general
nature:
\begin{itemize}
\item The Boltzmann H-theorem, namely the
definition of the entropy function to prove dissipation
\begin{equation}
S_B(t) = - \int_{{\mathbb R}^n} d {\bf x} \int_{\mathcal D}
p({\bf x},t;{\bf b}) \, \log p({\bf x},t;{\bf b}) \, d {\bf b}
\label{eq_8b_3}
\end{equation}
corresponds,   in the CGPK formalism to the
Boltzmann-Shannon entropy introduced in part II 
in the discrete $N$-state case, and here generalized
to a continuum of stochastic states. The entropy
function (\ref{eq_8b_3}) is  based on the
whole structure of the partial probability waves $p({\bf x},t;{\bf b})$,
and not on the overall density function $\int_{\mathcal D} p({\bf x},t;{\bf b})
\, d {\bf b}$.
\item The stochastic formulation permits to identify, from 
the transition kernel $K({\bf b},{\bf b}^\prime,[p])$, the
transition rate function
\begin{equation}
\lambda({\bf b};[p])=  \int_{{\mathcal D}} 
\int_{{\mathcal D}} \int_{{\mathcal D}}
H({\bf b}^\prime,{\bf b}^{\prime \prime \prime} | {\bf b}, 
{\bf b}^{\prime \prime }) \,
p({\bf x},t;{\bf b}^{\prime  \prime }) \, d {\bf b}^\prime
\, d {\bf b}^{\prime \prime} \,  d {\bf b}^{\prime \prime \prime }
\label{eq_8b_4}
\end{equation}
and the transition probability kernel
\begin{equation}
\hspace{-1.5cm}
A({\bf b},{\bf b}^\prime;[p]) = \frac{1}{\lambda({\bf b}^\prime;[p])}
\, \int_{{\mathcal D}} \int_{{\mathcal D}}
 H({\bf b}, {\bf b}^{\prime \prime \prime} |{\bf b}^\prime,{\bf b}^{\prime  \prime}) \,
p({\bf x},t,{\bf b}^{\prime \prime}) \, d {\bf b}^{\prime \prime} 
\, d {\bf b}^{\prime \prime \prime}
\label{eq_8b_5}
\end{equation}
which is a left-stochastic kernel. These quantities can be
useful in the mathematical analysis of the model
to assess the relaxation properties and the {\em propagation
of chaos} - to quote an expression by M. Kac \cite{kac_kinetic} - in the system.
\item The quantities $\lambda({\bf b};[p])$ and $A({\bf b},{\bf b}^\prime;[p])$
define completely a stochastic simulator of the Nonlinear CGPK process
associated with the Boltzmann equation, which can be viewed as 
 a stochastic molecular simulator.
\item The derivation of the nonlinear collisional Boltzmann equation
from a simple stochastic model justifies, on stochastic ground,
the intrinsic irreversibility associated with the
Boltzmann equation. In this framework, the irreversible behavior
is not related with the underlying, possibly chaotic,
 conservative Hamiltonian dynamics associated with the collision
process. This claim, 
is strongly related to
 the classical Zermelo's objection on the purely mechanical
interpretation of the Boltzmann equation based on the application of 
the Poincar\'e recurrence theorem, and the CGPK theory of the
kinetic equation, outlined above,  gives a simple answer to
the Zermelo's criticism.
In eqs. (\ref{eq_8b_1})-(\ref{eq_8b_2}), 
Hamiltonian conservative mechanics enters 
solely in order to specify the functional form of the
kernel $H({\bf b},{\bf b}^{\prime \prime \prime} | {\bf b}^\prime,
{\bf b}^{\prime \prime})$
in order to be consistent with the conservation
requirements (as regards momentum and kinetic energy)
dictated by the assumption of elastic collisions eq (\ref{eq_8b_agg2}).
The assumption of a rarefied particle-gas systems
enters in the linearity of $K({\bf b},{\bf b}^\prime;[p])$
with respect to the partial density waves $p({\bf x},t;{\bf b})$.
As the theory of CGPK processes concerns,
 the functional form of the
Boltzmann equation resides exclusively on the Markovian
character of the recombination mechanism  amongst the partial probability
waves.
\item The above derivation opens up relevant
issues in atomic physics associated with  stochasticity
and its physical meaning. The problem can be stated as follows:
given the above  derivation of the Boltzmann equation
starting from a purely stochastic model,  
the stochastic nature of a particle gas system is  solely a  mathematical
result following from CGPK theory or, rather it is a manifestation of
more fundamental processes underlying relaxation and irreversibility
in atomic and molecular systems? We believe that
the second approach would prove to be  correct, and further 
investigation would  hopefully lead to
new results on an old  classical subject,
in order to explain the dissipative behavior
of an ensemble of identical gas molecules mutually 
interacting via binary collisions.
\item As for the elastic Boltzmann equation, the
kinetic equations for granular materials  \cite{kinetic_granular},
for  which the
collisions are no longer 
 elastic, and an inelastic restitution coefficient
is introduced, can be treated on stochastic grounds using Nonlinear CGPK.
Similarly, it would be possible to identify stochastic models for
other kinetic equations (such as those of plasma physics) using CGPK processes.
In this respect, solely the collisionless Vlasov equation \cite{balescu}
represents an exception in this stochastic paradigm.  This
is not surprising, as the Vlasov equation is an isentropic model
that cannot be treated within a stochastic theory in which the
entropy production occurs as a consequence of the recombination
amongst the partial probability waves.
\end{itemize}

\section{Transport equations from GPK processes}
\label{sec_4}

The equations for the partial probability
waves represent the basic arche\-type of transport equations derived
from GPK processes.
Consistently with the principle of primitive variables, these
equations are parametrized with respect to the state of
the stochastic perturbation, expressed by the  partial probability densities
$p_\alpha({\bf x},t)$.
This represents a major difference with respect to 
transport equations derived starting from  Langevin
microdynamics driven by Wiener processes.
In the latter case, due to the independence of the increments,
the state of the stochastic perturbation is
completely renormalized out of the associated
Fokker-Planck equation.
In this Section, we analyze in some detail the functional
structure of the transport equations deriving from
GPK models. To begin with, we  develop the transport equation
from GPK dynamics associated with the evolution
of  the mean magnetic field in a solenoidal flow field (dynamo
problem) \cite{dynamo1,dynamo2}. Subsequently,  transport equations
for matter and momentum density in a fluid continuum are
derived starting from a GPK Ornstein-Uhlenbeck process.
Finally, the theory is extended to chemical reaction
kinetics.

\subsection{Transport of magnetic field and the GPK dynamo
problem}
\label{sec_4_1}

Let ${\bf v}({\bf x},t)$ be a solenoidal time-dependent velocity
field in ${\mathbb R}^n$,
and  consider a GPK process
$(n,N,{\boldsymbol \Lambda},{\bf A}, \{{\bf b}_\alpha\}_{\alpha=1}^N,{\bf v}({\bf x},t))$
 admitting an isotropic Kac limit, characterized
by the effective diffusivity $D_{\rm eff}$.
Let ${\bf B}({\bf x},t)$ be the magnetic field, and consider
the transport of ${\bf B}({\bf x},t)$ due to the  advective
action of the velocity field
${\bf v}({\bf x},t)$ in the case stochastic fluctuations are
superimposed. This is the essence of the dynamo
problem in the presence of diffusion admitting interesting astrophysical
applications \cite{astro1,astro2,astro3}

A stochastic microdynamic equation for this
process, written in the
form of a GPK process, can be expressed as
\begin{eqnarray}
d  {\bf x}(t) & = & {\bf v}({\bf x}(t)) d t + {\bf b}_{\chi_N(t)} \, dt
\nonumber \\
d {\bf B}(t) & = & {\bf G}({\bf x}(t), t) \cdot {\bf B}(t) \, d t
\label{eq10_1_1}
\end{eqnarray}
where ${\bf G}({\bf x}(t), t) \cdot {\bf B}(t)= \nabla {\bf v}({\bf x},t) \cdot
{\bf B}(t)$ accounts for the stretching of the magnetic
field induced by the velocity field ${\bf v}({\bf x}(t),t)$,
the $h$-entry of which is given by
\begin{equation}
\left ( {\bf G} \cdot {\bf B} \right )_h =
\left ( \nabla {\bf v} \cdot B \right )_h
= \sum_{k=1}^n \frac{\partial v_h}{\partial x_k} \, B_k
\label{eq10_1_2}
\end{equation}
Equation (\ref{eq10_1_1}) represents the stochastic formulation
of the magnetic dynamo problem under the assumption
that the magnetic field does not influence the evolution of
the flow field ${\bf v}({\bf x},t)$, corresponding
to the one-way coupling approximation.

The statistical description of this GPK process
involves the partial probability waves $p_\alpha({\bf x}, {\bf B}, t)$,
that are solution of the system of hyperbolic equations
\begin{eqnarray}
\hspace{-1.5cm}
\partial_t p_\alpha  =  - \nabla_x \cdot \left  (
{\bf v} \, p_\alpha \right ) -   \nabla_x  \cdot \left ({\bf b}_\alpha \,
 p_\alpha \right )
- \nabla_B \cdot  \left ( {\bf G} \cdot {\bf B} \, p_\alpha \right )
 -\lambda_\alpha \, p_\alpha + \sum_{\gamma=1}^N \lambda_\gamma A_{\alpha,\gamma} \, p_\gamma
\label{eq10_1_3}
\end{eqnarray}
where $\nabla_x$ and $\nabla_B$ are the nabla-operators
with respect to the ${\bf x}$- and ${\bf B}$-variables,
respectively.

The basic macroscopic observable in  the dynamo problem
is the mean magnetic field $\langle {\bf B}({\bf x},t) \rangle$
defined as
\begin{equation}
\langle {\bf B}({\bf x},t) \rangle = 
\int_{{\mathbb R}^n} {\bf B} \, p({\bf x},{\bf B},t) \,
d {\bf B}=\sum_{\alpha=1}^N \int_{{\mathbb R}^n}
{\bf B} \, p_\alpha({\bf x},{\bf B}, t) \, d {\bf B}
\label{eq10_1_4}
\end{equation}
and depending on the position ${\bf x}$ and time $t$.

The parametrization with respect to the stochastic state $\alpha$ suggests
to introduce the auxiliary quantities
\begin{equation}
\langle {\bf B}_\alpha({\bf x},t) \rangle 
=  \int_{{\mathbb R}^n}
{\bf B} \, p_\alpha({\bf x},{\bf B}, t) \, d {\bf B}
\label{eq10_1_5}
\end{equation}
Obviously,
\begin{equation}
\langle {\bf B}({\bf x},t) \rangle= \sum_{\alpha=1}^N
\langle {\bf B}_\alpha({\bf x},t) \rangle
\label{eq10_1_6}
\end{equation}
From eq. (\ref{eq10_1_3}), after some algebra,
the evolution equation for $\langle {\bf B}_\alpha({\bf x},t) \rangle$
can be derived. Componentwise, it reads
\begin{eqnarray}
\partial_t \langle B_{\alpha,h} \rangle
 &= & -{\bf v} \cdot \nabla_x  \langle
B_{\alpha,h} \rangle + \sum_{k=1}^n \frac{\partial v_h}{\partial x_k}
\, \langle B_{\alpha,k} \rangle  - {\bf b}_\alpha \cdot
\nabla_x \langle B_{\alpha,h} \rangle \nonumber \\
&- & \lambda_\alpha \, \langle B_{\alpha,h} \rangle +
\sum_{\gamma=1}^N \lambda_\gamma \, A_{\alpha,\gamma}
\, \langle B_{\gamma,h} \rangle
\label{eq10_1_7}
\end{eqnarray}
where $\langle B_{\alpha,h} \rangle$ is the $h$-entry
of $\langle {\bf B}_\alpha \rangle$, $h=1,\dots,n$.

In the Kac limit, this system of equations   
converges towards the solution
of the parabolic
equation
\begin{equation}
\partial_t \langle  {\bf B} \rangle
= - {\bf v} \cdot \nabla_x  \langle {\bf B} \rangle
+ \nabla_x {\bf v} \cdot \langle {\bf B} \rangle +
D_{\rm eff} \, \nabla_x^2 \langle {\bf B} \rangle
\label{eq10_1_8}
\end{equation}
involving the overall magnetic field
$\langle {\bf B} \rangle$ defined by eq. (\ref{eq10_1_6}),
which is the classical evolution equation
for the  dynamo problem in the presence of diffusion \cite{dynamo1,dynamo2}.

The system of equations (\ref{eq10_1_7}) represents
the exact transport equation for the
first-order partial moments $\langle {\bf B}_\alpha \rangle$
within the framework of GPK theory. 
The derivation of these equations does not
involve any constitutive assumption, and this is the
reason why we  have considered this problem as the first 
example of undulatory 
transport theory.

Maxwell equations dictates that the magnetic field should
be solenoidal. From eq. (\ref{eq10_1_7}) it follows
that 
\begin{eqnarray}
\hspace{-2.1cm}
\partial_t \nabla_x \cdot \langle {\bf B}_\alpha \rangle 
 =  - ({\bf v}+{\bf b}_\alpha) \cdot \nabla_x \left (
\nabla_x \cdot \langle {\bf B}_\alpha \rangle \right ) 
  -\lambda_\alpha \,  
\nabla_x \cdot \langle {\bf B}_\alpha \rangle 
+ \sum_{\gamma=1}^N \lambda_\gamma \, A_{\alpha,\gamma} \,
\nabla_x \cdot \langle {\bf B}_\gamma \rangle 
\label{eq10_1_9}
\end{eqnarray}
where the solenoidal nature of the velocity
field ${\bf v}({\bf x},t)$ has been enforced.
Equation (\ref{eq10_1_9}) indicates that,
 if all the partial averages of $\langle {\bf B}_\alpha({\bf x},0)
\rangle$, $\alpha=1,\dots,N$  at time $t=0$ are solenoidal,
i.e., $\nabla_x \cdot \langle {\bf B}({\bf x},0) \rangle=0$,
then the transport model (\ref{eq10_1_7}) 
preserves this property, namely $\nabla_x \cdot \langle {\bf B}_\alpha({\bf x},t) \rangle=0$, and {\em a fortiori},
$\nabla_x \cdot \langle {\bf B}({\bf x},t) \rangle=0$
for any time $t>0$.

The classical
 theory of the dynamo problem in the presence of diffusion
provides that, for  two-dimensional  spatial problems
(${\bf x}=(x_1,x_2) \in{\mathbb R}^2$), the $L^2$-norm of the magnetic field, solution
of the parabolic equation (\ref{eq10_1_8}),
decays exponentially in time  for generic smooth time-periodic velocity 
fields \cite{dynamo1}. Conversely, a positive dynamo action,
 i.e., an exponential
divergence with time  of the $L^2$-norm of the magnetic field may occur
starting from $n=3$.

It is easy to check that for $n=2$, in the case the average
magnetic field possesses zero mean, this property
holds also for the GPK dynamo equation (\ref{eq10_1_9}).
This result, stems straightforwardly from the
observation that the partial fields $\langle {\bf B}_{\alpha}({\bf x},t)
 \rangle= \left ( \langle B_{\alpha,1}({\bf x},t) \rangle,  \langle B_{\alpha,2}({\bf x},t) \rangle \right )$ can be expressed in terms of
a family of (scalar) vector potentials $\psi_\alpha({\bf x},t)$
\begin{equation}
\langle B_{\alpha,1}({\bf x},t) \rangle = \frac{\partial \psi_\alpha({\bf x},t)}{\partial x_2} \, , \qquad
\langle B_{\alpha,1}({\bf x},t) \rangle = -\frac{\partial \psi_\alpha({\bf x},t)}{\partial x_1}  
\label{eq10_1_10}
\end{equation}
$\alpha=1,\dots,N$
which, in turn, are solutions of the associated
advection-diffusion equations for a scalar field
\begin{eqnarray}
\partial_t \psi_\alpha & = & - {\bf v} \cdot \nabla_x \psi_\alpha
- {\bf b}_\alpha \cdot \nabla_x \psi_\alpha 
 - \lambda_\alpha \, \psi_\alpha + \sum_{\gamma=1}^N \lambda_\gamma
A_{\alpha,\gamma} \, \psi_\gamma
\label{eq10_1_11}
\end{eqnarray}
As shown in part II, GPK advection-diffusion of a scalar field
in the standard-map flow, the $L^2$-norms of $\psi_\alpha({\bf x},t)$
 decay exponentially to zero as a function of time.

Therefore, the effect of GPK perturbations is
essentially to modify the decay exponent (in two-dimensional
spatial problems) with respect to the  Kac limit (as
in the case of the chaotic advection-diffusion problem
for a scalar field addressed in part II),
but not the quality of stability. The three dimensional
case is fully open for investigation, and there is the
possibility that Poissonian perturbations could
modify  the stability properties of the diffusive dynamo problem, 
determining the
occurrence of a positive dynamo action, also in those
cases where the corresponding parabolic model 
(\ref{eq10_1_8}) possesses all the eigenvalues with negative
real part. The comparison with the analysis
developed by  Arnold and Korkina  \cite{korkina} and by Galloway and Proctor
\cite{proctor} for
the ABC flow would be an interesting benchmark of
this hypothesis.

\subsection{Mass and momentum transport}
\label{eq_4_2}

In this paragraph we consider the structure of mass and momentum
transport equation in a moving continuum as it emerges from GPK theory.
A moving continuum is nothing but an extremely useful macroscopic
approximation of the granularity of matter at microscale, 
resulting from the averaging of the local stochastic motion.

According  with the basic principles outlined in Section 2 of part I 
(principle of stochastic reality), let us consider for the
granular entities (be them  particles, molecules, aggregates, clusters,
 etc.), forming the continuous fluid phase, a GPK equation of
motion of the form
\begin{eqnarray}
d {\bf x}(t) & = & {\bf v}(t) \, dt \nonumber \\
m d {\bf v}(t) & = & {\bf f}({\bf x}(t)) \, dt +
{\bf b}_{\chi_N(t)} \, dt
\label{eq_10_2_1}
\end{eqnarray}
where ${\bf x} \in {\mathbb R}^n$, ${\bf v} \in {\mathbb R}^n$.
Equation (\ref{eq_10_2_1}) represents a Ornstein-Uhlenbeck
process, in which  all the microscopic ``granules'' possess
equal mass $m$, in the presence of a force field ${\bf f}({\bf x})$
and of stochastic fluctuations of Poisson-Kac nature,
described by a $N$-state finite Poisson process modulating 
a system of $N$ stochastic acceleration vectors 
${\bf b}_\alpha$, $\alpha=1,\dots,N$.  

Let $p_\alpha({\bf x},{\bf v},t)$, $\alpha=1,\dots,N$, be  the partial
probability densities associated with  eq. (\ref{eq_10_2_1}), which
satisfy the balance equations
\begin{eqnarray}
\hspace{-1.5cm}
\partial_t p_\alpha  =  - \nabla_x \cdot \left ( 
{\bf v} \, p_\alpha \right ) - \frac{1}{m} \, \nabla_v \cdot
\left ( {\bf f} \, p_\alpha  \right ) - \frac{1}{m} \, {\bf b}_\alpha \cdot
\nabla_v p_\alpha 
 -  \lambda_\alpha \, p_\alpha + \sum_{\gamma=1}^N \lambda_\gamma \, A_{\alpha,\gamma} \, p_\gamma
\label{eq_10_2_2}
\end{eqnarray}
where, as above, $\nabla_x$ and $\nabla_v$ represent the
nabla operators with respect to the ${\bf x}$- and ${\bf v}$-variables,
respectively. 
As in the classical setting of the hydrodynamic limit
from kinetic schemes, we are  interested in the lower-order moments
of the partial densities $p_\alpha({\bf x},{\bf v},t)$.
In the present analysis we focus exclusively on the
mass and momentum densities. A complete analysis
involving energy density will be developed in a forthcoming work.

Let
\begin{equation}
\rho_\alpha({\bf x},t) = m \, \int_{{\mathbb R}^n} p_\alpha({\bf x},{\bf v},t)
\, d {\bf v}
\label{eq_10_2_3}
\end{equation}
\begin{equation}
\rho_\alpha({\bf x},t) \, {\bf V}_\alpha({\bf x},t)
= m \, \int_{{\mathbb R}^n} {\bf v} \,  p_\alpha({\bf x},{\bf v},t)
\, d {\bf v}
\label{eq_10_2_4}
\end{equation}
$\alpha=1,\dots,N$, be  the partial mass ($\rho_\alpha$)
and momentum ($\rho_\alpha \, {\bf V}_\alpha$) densities,
respectively.
The overall mass ($\rho({\bf x},t)$) and momentum ($\rho({\bf x},t)
\, {\bf V}({\bf x},t)$) densities
are the sum with respect to $\alpha$ of the  corresponding
partial quantities
\begin{equation}
\rho({\bf x},t) = \sum_{\alpha=1}^N \rho_\alpha({\bf x},t)
\, , \qquad
\rho({\bf x},t) \, {\bf V}({\bf x},t)
= \sum_{\alpha=1}^N \rho_\alpha({\bf x},t)
\, {\bf  V}_\alpha({\bf x},t)
\label{eq_10_2_5}
\end{equation}
From eq.  (\ref{eq_10_2_2}) it follows that $\rho_\alpha({\bf x},t)$
satisfy the system of partial continuity equations
\begin{equation}
\partial_t \rho_\alpha = - \nabla_x \cdot \left ( \rho_\alpha \, {\bf V}_\alpha
\right ) - \lambda_\alpha \, \rho_\alpha + \sum_{\gamma=1}^N 
\lambda_\gamma \, A_{\alpha,\gamma} \, \rho_\gamma 
\label{eq_10_2_6}
\end{equation}
that, once summed over $\alpha$, provide the overall
continuity equation
\begin{equation}
\partial_t \rho({\bf x},t) = - \nabla_x \cdot \left (
\rho({\bf x},t) \, {\bf V}({\bf x},t) \right ) 
\label{eq_10_2_7}
\end{equation}
Next, consider the partial momentum densities. Multiplying eq. 
(\ref{eq_10_2_2}) by $m \, {\bf v}$ and integrating over ${\bf v}$
one  obtains
\begin{eqnarray}
\partial_t \left ( \rho_\alpha \, {\bf V}_\alpha \right )
&  = &  - \nabla_x \cdot \left ( \int_{{\mathbb R}^n} 
m \, {\bf v} \, {\bf v} \, p_\alpha \, d {\bf v} \right )
+ \frac{{\bf f}}{m} \, \rho_\alpha + \frac{{\bf b}_\alpha}{m}
\, \rho_\alpha  \nonumber \\
& - & \lambda_\alpha \, \left (\rho_\alpha \, {\bf V}_\alpha \right )
+ \sum_{\gamma=1}^N \lambda_\gamma \, A_{\alpha,\gamma}
\, \left ( \rho_\gamma \, {\bf V}_\gamma \right )
\label{eq_10_2_8}
\end{eqnarray}
Let ${\bf T}_\alpha({\bf x},t)$ be the comprehensive partial stress tensor,
corresponding to the dyadic second-order term
\begin{equation}
{\bf T}_\alpha({\bf x},t) =  m  \, \int_{{\mathbb R}^n} 
 {\bf v} \, {\bf v} \, p_\alpha({\bf x},{\bf v},t) \, d {\bf v}
\label{eq_10_2_9}
\end{equation}
$\alpha=1,\dots,N$, that includes also the inertial contribution.
The comprehensive partial stress tensor can be dissected into
a partial inertial contribution $\rho_\alpha \, {\bf V}_\alpha \, {\bf V}_\alpha$,
and into a partial stress tensor ({\em sensu stricto}) $\widehat{\boldsymbol
\tau}_\alpha$,
\begin{equation}
{\bf T}_\alpha = \rho_\alpha \, {\bf V}_\alpha \, {\bf V}_\alpha
+ \widehat{\boldsymbol \tau}_\alpha
\label{eq_10_2_10}
\end{equation}
which in turn can be developed, as in classical continuum
mechanics  \cite{de_groot},
 into a traceless stress tensor ${\boldsymbol \tau}_\alpha$ and into a compressive isotropic pressure contribution $P_\alpha \,{\bf I}$ (${\bf I}$ is the identity tensor)
\begin{equation}
\widehat{\boldsymbol \tau}_\alpha = {\boldsymbol \tau}_\alpha
+ P_\alpha \, {\bf I}
\label{eq_10_2_11}
\end{equation}
$\alpha=1,\dots,N$, where $\mbox{Trace}({\boldsymbol \tau})=0$.
Enforcing the above decompositions,
one obtains the system of momentum balance
equations
\begin{eqnarray}
\partial_t \left ( \rho_\alpha \, {\bf V}_\alpha \right )
& = & - \nabla_x \cdot  \left (\rho_\alpha \, {\bf V}_\alpha \, {\bf V}_\alpha \right )
- \nabla_x \cdot {\boldsymbol \tau}_\alpha - \nabla_x P_\alpha
+ \frac{{\bf f}}{m} \, \rho_\alpha  \nonumber \\
&+& \frac{{\bf b}_\alpha}{m} \, \rho_\alpha
- \lambda_\alpha \, \left ( \rho_\alpha \, {\bf V}_\alpha \right )
+ \sum_{\gamma=1}^N \lambda_\gamma \, A_{\alpha,\gamma} \, \left ( \rho_\gamma \, {\bf V}_\gamma \right )
\label{eq_10_2_12}
\end{eqnarray}
$\alpha=1,\dots,N$, that represent the general setting of momentum
transfer in the hydrodynamic limit of GPK theory  for a 
single phase/single component moving continuum.

Summing eq. (\ref{eq_10_2_12}) over $\alpha$, the
balance equation for the overall momentum
density is obtained
\begin{eqnarray}
\partial_t \left ( \rho \, {\bf V} \right )
&= & - \nabla_x \cdot \left ( \sum_{\alpha=1}^N \rho_\alpha \, {\bf V}_\alpha
\, {\bf V}_\alpha \right )
- \nabla_x \cdot {\boldsymbol \tau} - \nabla_x P + \frac{ \rho \, {\bf f}}{m} 
\nonumber \\
&+ & \sum_{\alpha=1}^N \frac{\rho_\alpha \, {\bf b}_\alpha}{m}
\label{eq_10_2_13}
\end{eqnarray} 
where the overall stress tensor ${\boldsymbol \tau}$ and pressure
$P$ are simply the sum over $\alpha$ of the corresponding partial
quantities
\begin{equation}
{\boldsymbol \tau}= \sum_{\alpha=1}^N {\boldsymbol \tau}_\alpha \, ,
\qquad P= \sum_{\alpha=1}^N P_\alpha
\label{eq_10_2_14}
\end{equation}

There are two main qualitative differences with respect to the
classical hydrodynamic formulation of momentum
transport:
\begin{itemize}
\item the inertial term does not reduce
to $\rho \, {\bf V} \, {\bf V}$, but contains explicit
reference (memory) of all the partial inertial contributions
expressed by the term $\sum_{\alpha=1}^N \rho_\alpha \, {\bf V}_\alpha
\, {\bf V}_\alpha$;
\item there is an additional contribution expressed by
the term $\sum_{\alpha=1}^N \rho_\alpha \, {\bf b}_\alpha /m$ accounting
for the nonuniformity effects of the stochastic acceleration terms
amongst the partial structures of the fluid mixture.
This term obviously vanishes in the Kac limit where, due to
the fast recombination amongst the partial probability
 waves $\rho_\alpha \simeq \rho/N$
and $\sum_{\alpha=1}^N \rho_\alpha \, {\bf b}_\alpha/m \simeq 0$,
due to the zero-bias condition $\sum_{\alpha=1}^N {\bf b}_\alpha=0$.
\end{itemize}
It follows from the above observations, that in the GPK theory
of hydrodynamics one cannot reduce mass and momentum transport exclusively
to the analysis of the overall fields $\rho({\bf x},t)$ and 
${\bf V}({\bf x},t)$,
but one is forced to solve simultaneously mass and momentum
balance equations for the full system of partial
field \{$\rho_\alpha({\bf x},t)$, ${\bf V}_\alpha({\bf x},t) \}_{\alpha=1}^N$.

Also for the GPK mass and momentum balance equations, the concept
of Kac limit applies, and these equations should reduce
for $b^{(c)}, \lambda^{(c)} \rightarrow \infty$, keeping constant
the nominal diffusivity, to the usual continuity
and Navier-Stokes equations, upon a suitable assumption
on the constitutive equations for ${\boldsymbol \tau}_\alpha({\bf x},t)$.

A discussion on the constitutive equations for  the
partial stress tensors ${\boldsymbol \tau}_\alpha({\bf x},t)$, as well as
a comprehensive analysis of the GPK hydrodynamics and its qualitative
differences with respect to the classical approach will be developed
in  a forthcoming article.

\subsection{Chemical reactions}
\label{sec_5_3}

In this paragraph we briefly  discuss the modeling of chemical reaction
kinetics within the framework of GPK theory. Consider the simplest
case of a  bimolecular elementary isothermal chemical reaction
\begin{equation}
A + B \rightarrow P
\label{eq_10_3_1}
\end{equation}
 in ${\mathbb R}^n$, 
the rate of
which, in the mean-field limit, is given by
\begin{equation}
r(c_A,c_B) = k_r \, c_A \, c_B
\label{eq_10_3_2}
\end{equation}
where $c_A$, $c_B$ are the molar concentrations of the two
reacting species, and $k_r$ is the rate coefficient independent 
of the concentrations but eventually function of temperature.
Assume that the reacting process  evolves in a fluid continuum
where the reacting molecules of the two species
 are subjected to a deterministic
drift ${\bf v}({\bf x})$ and to stochastic fluctuations
expressed by means of a GPK process.

Consider a GPK process possessing a finite number
$N$ of states. 
Let $c_{A,\alpha}({\bf x},t)$, $c_{B,\beta}({\bf x},t)$
be the partial (molar) concentrations of the two reacting
species parametrized with respect to  the stochastic
  state $\alpha=1,\dots,N$.
The overall molar concentrations are just the sum of the
partial concentrations with respect to the stochastic index $\alpha$
\begin{equation}
c_A({\bf x},t)= \sum_{\alpha=1}^N c_{A,\alpha}({\bf x},t)
\, , \qquad
c_B({\bf x},t)= \sum_{\alpha=1}^N c_{B,\alpha}({\bf x},t)
\label{eq_10_3_3}
\end{equation}
Let $(n,N,{\boldsymbol \Lambda},{\bf A}, \{{\bf b}_\alpha \}_{\alpha=1}^N,
{\bf v}({\bf x}))$ be  a GPK process a admitting in the
Kac limit an effective diffusivity $D_{\rm eff}$.
Taking into account the presence of the deterministic velocity field
${\bf v}({\bf x})$, the GPK balance equations for the
partial concentrations of the two reacting species can be
expressed by
\begin{eqnarray}
\hspace{-2.2cm}
\partial_t c_{A,\alpha}    =   - \nabla \cdot \left ( {\bf v} \, c_{A,\alpha}
\right ) - {\bf b}_\alpha \cdot \nabla c_{A,\alpha}
- k_r \, \left ( \sum_{\gamma=1}^N c_{B,\gamma} \right ) \, c_{A,\alpha}
 -\lambda_{\alpha} \, c_{A,\alpha} + \sum_{\gamma=1}^N
\lambda_\gamma \, A_{\alpha,\gamma} \, c_{A,\gamma}
\nonumber \\
\hspace{-2.2cm}
\partial_t c_{B,\alpha}  =  - \nabla \cdot \left ( {\bf v} \, c_{B,\alpha}
\right ) - {\bf b}_\alpha \cdot \nabla c_{B,\alpha}
- k_r \, \left ( \sum_{\gamma=1}^N c_{A,\gamma} \right ) \, c_{B,\alpha}
  -\lambda_{\alpha} \, c_{B,\alpha} + \sum_{\gamma=1}^N
\lambda_\gamma \, A_{\alpha,\gamma} \, c_{B,\gamma} \nonumber 
\\
\label{eq_10_3_5}
\end{eqnarray}
$\alpha=1,\dots,N$.
Observe that the way a chemical reaction can be  introduced
within the GPK balance equation is not unique.
The two reacting contributions $k_r \left (\sum_{\gamma=1}^N c_{B,\gamma}
\right )  \, c_{A,\alpha}$, $k_r \left (\sum_{\gamma=1}^N c_{A,\gamma}
\right )  \, c_{B,\alpha}$, entering the balance equations
for $c_{A,\alpha}$ and $c_{B,\alpha}$, respectively correspond to
the assumption that the reaction rate at  the space-time point 
$({\bf x},t)$
 depends exclusively
on the actual overall concentrations of the two reacting species at $({\bf x},t)$. Other choices are also possible.

In the Kac limit, eqs. (\ref{eq_10_3_5}) converge towards 
the system of two parabolic equations for the overall
concentrations $c_A({\bf x},t)$ and $c_B({\bf x},t)$,
\begin{equation}
\partial_t c_h = - \nabla \cdot \left ( {\bf v} \, c_h \right )
+ D_{\rm eff} \nabla^2 c_h - k_r c_A c_B \,, \;\;\; h=A,B
\label{eq_10_3_6}
\end{equation}

In point of fact, the formal structure of Continuum GPK processes
provides a natural way to account for the collision
efficiency and its  influence on the reaction rate, in an analogous
way  the reactive Boltzmann equation does \cite{reactive_boltzmann}. This is
by no mean surprising, due to the equivalence between CGPK process 
in the presence of NFPK kernels and the non-reactive
Boltzmann equation developed  in Section \ref{sec_3}.
Next,  we consider briefly this class of models, in the
absence of a deterministic velocity field ${\bf v}({\bf x})=0$ (for
simplifying the notation),
assuming a linear model where both  ${\boldsymbol \Lambda}$
and ${\bf A}$ do not depend on the partial density waves.
The state of the stochastic 
perturbation is parametrized
with respect to the stochastic velocity vector ${\bf b}$, attaining values
in ${\mathcal D} \subseteq {\mathbb R}^n$,
and the bimolecular reaction (\ref{eq_10_3_1}) is considered
under isothermal conditions.

Consequently, 
the partial concentrations $c_A({\bf x},t;{\bf b})$, $c_B({\bf x},t;{\bf b})$
in this setting depend continuously on the stochastic
velocity vector ${\bf b}$.
As regards the chemical kinetic contribution to the evolution
of $c_A({\bf x},t,{\bf b})$, $c_B({\bf x},t,{\bf b})$,
two reaction kernels $R_A({\bf b},{\bf b}^\prime)$, $R_B({\bf b},{\bf b}^\prime)$ are introduced, representing the fraction
 per unit time of colliding 
molecules  of $A$ and $B$, respectively,
possessing stochastic velocities ${\bf b}$ and ${\bf b}^\prime$
which  perform the reaction, i.e.,  that are able to overcome the
reaction activation energy. Although not explicited, these
kernels depend on temperature by a Arrenhius-Kramers factor
$\exp(-E_R/k_B T)$, where $E_R$ is the reaction activation energy, $k_B$
the Boltzmann constant, and $T$ the absolute temperature.
Stoichiometry dictates that $R_A=R_B=R$ and, moreover
the reaction kernel can be assumed, for simplicity, symmetric with respect to
their argument, namely
\begin{equation}
R({\bf b},{\bf b}^\prime)= R({\bf b}^\prime,{\bf b})
\label{eq_10_3_7}
\end{equation}
Taking into account the analysis developed in Section
\ref{sec_3}, the balance 
equations for $c_A({\bf x},t,{\bf b})$, $c_B({\bf x},t,{\bf b})$
now read
\begin{eqnarray}
\hspace{-1.5cm}
\partial_t c_A  =  - {\bf b} \cdot \nabla_x c_A-
\int_{{\mathcal D}} K({\bf b},{\bf b}^\prime) \, \left [
c_A - c_A^\prime \right ] d {\bf b}^\prime - c_A \, \int_{{\mathcal D}}
R({\bf b},{\bf b}^\prime) \, c_B^\prime \, d {\bf b}^\prime
\nonumber \\
\hspace{-1.5cm}
\partial_t c_B  =  - {\bf b} \cdot \nabla_x c_B-
\int_{{\mathcal D}} K({\bf b},{\bf b}^\prime) \, \left [
c_B - c_B^\prime \right ] d {\bf b}^\prime - c_B \, \int_{{\mathcal D}}
R({\bf b},{\bf b}^\prime) \, c_A^\prime \, d {\bf b}^\prime
\label{eq_10_3_8}
\end{eqnarray}
where $c_A=c_A({\bf x},t;{\bf b})$, $c_B=c_B({\bf x},t;{\bf b})$,
$c_A^\prime=c_A({\bf x},t;{\bf b}^\prime)$, $c_B^\prime
=c_B({\bf x},t;{\bf b}^\prime)$, and $K({\bf b},{\bf b}^\prime)$
is the transition kernel of the Continuous GPK process.
Given a physically motivated expression for the reaction
kernel $R({\bf b},{\bf b}^\prime)$, depending on the
chemistry of the reactive step,
 the Kac limit of the model, or alternatively
the application of homogenization techniques in the long-term
regime, provide an expression for the effective 
reaction coefficient $k_r$, entering the mean-field
model (\ref{eq_10_3_2}). The flexibility
of GPK models, especially in their continuous
formulation, provides a direct connection between momentum
transfer and the collisional efficiency of   a reactive process.

In a similar way, it is possible to
derive stochastic models for particle/antiparticle
production/annihilation accounting for
the conservation laws  (parity, energy and momentum) of the process.
The main physical issue in this case is not the annihilation
contribution, which is substantially analogous
to a bimolecular reaction, but the production term which intrinsically
depends on the vacuum fluctuations and involves the stochastic
characterization of zero point energy. The latter issue 
can be properly formalized  in the present theory,
embedding CGPK processes within
the framework of the second quantization
of the electromagnetic field, and its stochastic characterization
\cite{milonni}.

\section{Concluding remarks}
\label{sec_5}

We have introduced the concept of Generalized Poisson-Kac processes,
analyzed
their structural properties (part I) and  addressed their physical 
implications in statistical physical, hydrodynamics and transport theory 
(parts II and III).

In its very essence, a GPK process stems from the original intuition of Marc
Kac of considering   dichotomous velocity fluctuations possessing
finite propagation velocity,  
and considers the fluctuating contribution
as a transition amongst a finite number $N$ of
stochastic states immersed in a Markovian structure accounting
for the transitions.
The primitive statistical description of this process
involves $N$ partial probability density functions
$\{p_\alpha({\bf x},t)\}_{\alpha=1}^N$
the spatio-temporal evolution of which follows
a 
 hyperbolic  dynamics, corresponding to
a planar wave-motion in the presence of recombination.
This is the reason why $p_\alpha({\bf x},t)$ are also
referred to as ``partial probability waves'', and the
resulting macroscopic processes indicated as ``undulatory
transport model'', just to mark the wave-like nature
of their basic statistical descriptors.

The nonlinear extension of the theory, as well as the generalization
to a continuum of states, provides the natural setting
for obtaining straightforwardly a stochastic derivation
of the kinetic Boltzmann equation. This result opens up
fundamental issues on the underlying physical nature
of this equation, alternative to the purely mechanical (Liouvillian)
picture. In some sense, the stochastic derivation of the
collisional  Boltzmann equation concludes the original
Kac's program  in kinetic theory starting from
the 1954 paper based on a Markovian toy model for gas dynamics, 
and gives to it  new  impetus
for a thorough exploitation of a fully stochastic formulation of the
Boltzmann collisional
dynamics.

The GPK approach towards the Boltzmann equation is 
consistent with the Zermelo objection  \cite{zermelo1,zermelo2}
against the fully
mechanical (Hamiltonian) derivation of the Boltzmann
equation which, if postulated, leads intrinsically to 
finite-time Poincar\'e recurrences and to the lack of a truly irreversible
behavior. Albeit sedated by the analysis of the order of magnitude
of the recurrence time \cite{boltzmann1,chandra}, Zermelo objection  persists,
from the
strict mathematical and logical point of view, 
as a bleeding wound on the connection between a mechanical
(conservative) view of dynamics and  the  intrinsically
irreversible nature of thermodynamics.
All these issues on the foundation of Boltzmann collision equation
will be hopefully developed in a subsequence communication.
But it is rather intuitive, that the results obtained on the connections
between kinetic theory and NFPK models admit powerful implications,
not only from a theoretical and thermodynamic point of view, but
also in practical applications. The idea of developing a fully
stochastic molecular simulator based on NFPK dynamics is not
only intriguing, but also feasible and potentially computationally
advantageous with respect to the existing methods. Further studies
will clarify and quantify the correctness of this possibility.

What is also interesting to observe is that the physical concept of
collision amongst molecules corresponds, in the Continuous
GPK setting, to a Markovian recombination amongst the partial
probability waves $p({\bf x},{\bf v},t)$, induced by the choice of
the velocity ${\bf v}$ as the stochastic vector-valued variable
parametrizing the states of the CGPK process.

Apart from this result, GPK theory provides  a simple
and tractable class of processes that overcome
the intrinsic problems of Wiener-driven stochastic
models in describing the physical reality  (infinite propagation
velocity), and  permits to derive
out of them  new classes of transport equations in the  hydrodynamic (continuum) limit.
The main qualitative difference between Wiener-driven and GPK stochastic
models resides in the
{\em regularity issue}: the trajectory of GPK processes,
for finite values of $b^{(c)}$ and $\lambda^{(c)}$, are, with
probability $1$, almost everywhere smooth curves
of time, possessing fractal dimension $d_f=1$, and local
H\"{o}lder exponent $1$. Out of it, finite propagation
velocity follows.
Moreover, the Kac-limit property provides the
natural connection between GPK theory and
the classical stochastic formulation
of microdynamics  based on Brownian motion  and  Wiener-driven Langevin equations.
This connection is not only related to the Kac limit
 for $b^{(c)}, \lambda^{(c)} \rightarrow \infty$,
keeping fixed the nominal diffusivity $D_{\rm nom}$, but also as an
emergent property in the long-time
regime. This justifies while Brownian features can be observed
at time-scales much larger than the characteristic recombination
scale $t_{\rm recomb}=1/\lambda^{(c)}$.
The asymptotic properties of GPK processes 
 represent a bridge between GPK theory
and the statistical physics based on Brownian motion  and Wiener-driven
processes that, as said,
can be regarded as an emergent feature for long timescales.

The regularity issue of GPK processes admits
several implications:
\begin{itemize}
\item GPK processes, once extended to space-time stochastic
perturbations, provide a valuable tool for  approaching, on a rigorous
but save  way (as singularity issues are concerned),  spatio-temporal stochastic
dynamics and field-theoretical models (i.e., SPDE) driven
by stochastic perturbations. We have outlined several simple
linear examples in part II, and the same approach can
be extended to work out nonlinear models such as Burgers' equations,
KPZ model, $\Phi^4$-field stochastic quantization, etc.
\item The finite propagation velocity of GPK provides
a safeguard in the extension of GPK processes to the relativistic
case \cite{giona_rel1,giona_rel2}.
\item Trajectory regularity, characterizing GPK processes,
substantially simplifies all the subtleties and technicalities
of stochastic (Ito, Strato\-novich, Klimontovich etc.) calculus, as reduces it
to the simplest possible form, namely that of Riemann-Stieltjes
integrals.
\end{itemize}

There is another issue, mentioned throughout this work, that
deserves further attention, as it marks a
conceptual difference between Wiener-driven Langevin equations
and the corresponding stochastic models driven by GPK
processes. Consider a Langevin equation
driven by a Wiener forcing on the state variable ${\bf x}(t)$.
The Wiener forcing expresses in a lumped, coarse-grained way, a 
manifold of small-scale perturbations resulting from the
microscopic interactions, internal to the system, and from the
interactions with the surrounding. As a result,
the associated forward Fokker-Planck equation defines the
evolution of the probability density function in which 
there is no further reference on the state of the stochastic perturbation:
this makes this  class of models strictly Markovian.

Conversely, in the corresponding GPK case, the statistical
description of the process still involves information about the stochastic
perturbation:
for this reason a system of partial probability densities,
$\{p_\alpha({\bf x},t) \}_{\alpha=1}^N$ in the discrete case,
$p({\bf x},t,\alpha)$, $\alpha \in {\mathcal D}$
in the CGPK case, is required to describe the 
statistical evolution of the system. The fact that the statistical
description of the process keeps memory of the 
state of the stochastic perturbation, so that system dynamics and stochastic
perturbations cannot be decoupled, other than in the
Kac limit, is the physical origin of the trajectory regularity
of GPK processes.
Consequently, GPK processes are not strictly Markovian,
which respect to the overall probability density function
$p({\bf x},t)$, but an extended Markovian property can be still
established with respect to the complete probabilistic
description accounting also for the state of the stochastic
perturbation \cite{giona_markovian}.

The consequence of this property, as regards  the transport equations
in the hydrodynamic limit, is evident: a system driven
by GPK fluctuations is fully described  in the hydrodynamic
limit by a family of partial mass and  momentum densities
$\{\rho_\alpha$, $\rho_\alpha \, {\bf V}_\alpha \}_{\alpha=1}^N$,
and similarly for the other thermodynamic quantities,
consistently with the retained information on the state of the
stochastic perturbation.
This, namely the partial-wave approach towards the coarse-graining of
the microdynamic equation of motion, will be analyzed in future
works. In any case, it represents a novel and promising alternative
to the existing higher-moment expansions, that starting from
the well-known Grad's 13-moment approach \cite{grad}, have been developed
in the current literature on kinetic theory and statistical mechanics.

All the higher-moment expansions, beyond the classical 5-mode approach
associated with the collisional invariants, suffer
the intrinsic {\em vulnus}, as  regards the lack of positivity,
that can be  easily interpreted within the  Pawula theorem \cite{pawula}.
In this framework, the use of GPK theory, and of its implications
in the definition of the hydrodynamic limit, represents
not only a new way for approaching the coarse-graining
of microscopic dynamics towards the hydrodynamic equations for
a continuum, but also
the stochastic background for the development
of a rigorous, stochastically
consistent, formulation of the Extended Thermodynamic
Theories of Irreversible Processes, initiated with
the works by M\"uller and Ruggieri, and subsequently
elaborated by Jou, Lebon, Casas-Vazquez and many others,
aimed at generalizing the classical de-Groot Mazur theory
of irreversible processes, by introducing a more general
  definition
of the thermodynamic state variables  out of equilibrium
in order to include
the  contribution of the
 fluxes.

\end{document}